\documentclass[traditabstract]{aa} 

\usepackage{graphicx}
\usepackage{rotating}
\usepackage{epsfig}
\usepackage{float} 
\usepackage{subfigure}
\usepackage[para,online,flushleft]{threeparttable}
\usepackage{natbib}
\bibpunct{(}{)}{;}{a}{}{,} 
\usepackage{dcolumn}

\usepackage[flushleft]{threeparttable}

\newcolumntype{.}{D{.}{.}{-1}}
\newcolumntype{;}{D{;}{.}{7}}

\usepackage{color}

\begin{document}

\title{First results from a large-scale proper motion study of the Galactic Centre}
\subtitle{}

\author{B. Shahzamanian\inst{1}, R. Sch\"odel\inst{1}, F. Nogueras-Lara\inst{1}, H. Dong\inst{1}, \\E. Gallego-Cano\inst{1,2}, A. T. Gallego-Calvente\inst{1}, A. Gardini\inst{1}}


   \institute{Instituto de Astrofísica de Andalucía (CSIC), Glorieta de la Astronomía s/n, 18008 Granada, Spain\\
       \email{shahzaman@iaa.es}
          \and
             Centro Astronómico Hispano-Alemán (CSIC-MPG),
             Observatorio Astronómico de Calar Alto, Sierra de los Filabres,
             04550, Gérgal, Almería, Spain
  }

\date{Received:/ Accepted:  }


\abstract {
Proper motion studies of stars in the centre of the Milky Way have been typically limited to the Arches and Quintuplet clusters and to the central parsec. Here, we present the first results of a large-scale proper motion study of stars within several tens of parsecs of Sagittarius A* based on our $0.2''$ angular resolution GALACTICNUCLEUS survey (epoch 2015) combined with NICMOS/HST data from the Paschen-$\alpha$ survey (epoch 2008). This study will be the first extensive proper motion study of the central $\sim 36' \times 16'$ of the Galaxy, which is not covered adequately by any of the existing astronomical surveys such as Gaia because of its extreme interstellar extinction ($A_{V} \gtrsim 30$ mag). Proper motions can help us to disentangle the different stellar populations along the line-of-sight and interpret their properties in combination with multi-wavelength photometry from GALACTICNUCLEUS and other sources. It also allows us to infer the dynamics and interrelationship between the different stellar components (Galactic bulge, nuclear stellar disk, nuclear stellar cluster) of the Galactic Centre (GC). In particular, we use proper motions to detect co-moving groups of stars which can trace low mass or partially dissolved young clusters in the GC that can hardly be discovered by any other means. Our pilot study in this work is on a field in the nuclear bulge associated by H{\scriptsize II} regions that show the presence of young stars. We detect the first group of co-moving stars coincident with an H{\scriptsize II} region. Using colour-magnitude diagrams, we infer that the co-moving stars are consistent with being the post-main sequence stars with ages of few Myrs. Simulations show that this group of stars is a real group that can indicate the existence of a dissolving or low to intermediate mass young cluster. A census of these undiscovered clusters will ultimately help us to constrain star formation at the GC in the past few ten Myrs.
}


\authorrunning{B. Shahzamanian} 
\titlerunning{Galactic Centre proper motions}
\maketitle
\section{Introduction}
\label{section:Introduction}

The Galactic Centre (GC) contains the nearest galactic nucleus at only 8 kpc from Earth that harbours the massive black hole (MBH), Sagittarius~A* (Sgr~A*) with a mass of $\sim 4 \times$ $10^{6}$ $\mathrm{M_{\odot}}$ \citep{Eckart&Genzel1996,Eckart&Genzel1997,Schoedel2003,Ghez2008, Gillessen2009stars,Boehle:2016zr, parsa2017,gravity2018}.

In the past $\sim$30 Myr, the GC has been the most active star forming region in the Milky Way on scales of $\sim$~100 pc \citep{Figer:2004fk, Matsunaga:2011uq, yusefzadeh2009}. The studies of current star formation in the GC have been concentrated on three known massive young clusters (the Arches, Quintuplet, and Central clusters) that lie within 30 pc of Sgr~A* \citep[ages $\sim$ 2-7 Myr, masses $\sim$ 3$\times 10^{4}$$\mathrm{M_{\odot}}$]{Figer:1999uq, Figer:2002qf, Launhardt:2002nx, Genzel:2003it, martins2008, Clarkson:2012fk, Schodel:2014fk, schodel:2014bn, stolte:2015, clark:2018, hosek:2015, hosek:2019,Gallego:2019}. For the star formation history in the few central parsecs of the Galaxy see also \cite{Blum:2003, Maness:2007, Pfuhl:2011uq}. Within 40 pc in projection of Sgr~A*, three classical Cepheids have been recently discovered that determine tight constraints on recent star formation in the nuclear stellar Disk (NSD), which is a disk-like stellar structure with the scale height of $\sim$45 pc and radius of 150-200 pc \citep{Launhardt:2002nx}, and imply a rate of $\sim$ 0.08 $\mathrm{M_{\odot}}$ yr$^{-1}$ in the past 30 Myr and a lower star formation rate $\gtrsim30$~Myr ago \citep[Fig.~3 in][]{Matsunaga:2011uq}. The NSD contains the nuclear star cluster (NSC), with a half-light radius of $\sim$5 pc and a mass of $\sim 2.5 \times$ $10^{7}$ $\mathrm{M_{\odot}}$ \citep{Schodel:2014fk}. NSCs are very common in all types of galaxies, are characterised by a complex population, and frequently co-exist with MBHs \citep{Neumayer:2017}. NSD and NSC are embedded in the nuclear bulge (NB), which is a stellar structure prominent from the kiloparsec-scale Galactic Bulge/Bar.

Due to the high recent star formation rate, the existence of at least ten times more undiscovered young ($\lesssim50$~Myr) massive clusters, or a larger number of clusters with smaller masses, than what are known in the nuclear disk is predicted \citep{Matsunaga:2011uq}. 
Young clusters can easily escape detection at the GC when they are either not very massive or dense or older than a few Myr. \cite{Portegies-Zwart:2002fk} have shown that even a cluster as massive as the Arches will be confused with the dense stellar background at the GC after a short time as 10 Myr.

The extreme and spatially highly variable interstellar extinction toward the GC \citep{Nishiyama:2008qa, Fritz:2011fk, Nogueras:2019b} makes it practically impossible to identify young clusters in the near-infrared (NIR) colour-magnitude diagrams (CMDs), which are almost fully degenerate because of reddening \citep{Nogueras2018a}. However, stellar proper motions can provide us with an alternative way of detecting young clusters, which are expected to show coherent proper motions. The internal velocity dispersion of the Quintuplet and Arches clusters ($\lesssim0.2 ~\mathrm{mas\, yr^{-1}}$ or $8~\mathrm{km\, s^{-1}}$) is significantly smaller than the velocity dispersion of the surrounding field stars \citep[on the order of $1-3~\mathrm{mas\, yr^{-1}}$ or $40 - 80~\mathrm{km\, s^{-1}}$, depending on the Gaussian mixture models adopted;][] {Stolte:2008uq, rui:2019, hosek:2019}.

Proper motion measurements of stars at the GC can also serve an important purpose in disentangling different overlapping Galactic components (e.g. Bulge and NSD) and thus infer their structures and formation histories on clean samples and derive the global structure and dynamics of the NSD \citep[][and references therein]{Matsunaga2018}. Studying proper motions may also answer the question whether the isolated massive stars found in the nuclear bulge are runaway stars from the known GC young massive clusters or whether they formed locally \citep{Wang:2010fk, Dong:2011ff}. It is currently generally accepted that massive stars always form as part of large clusters/groups. 
\\

This paper explores the potential of proper motion measurements at the GC. While one would ideally use homogeneously acquired data from the same instrument over a long time frame to measure proper motions, such data are not available for the large-scale GC environment. Instead, we use data from two high angular resolution surveys, the GALACTICNUCLEUS survey \citep[epoch 2015;][]{Nogueras2018prep} and the HST Paschen alpha survey \citep[epoch 2008;][]{Wang:2010fk, Dong:2011ff}. The purpose of this paper is two-fold: (1) Demonstrate the feasibility of our methodology by comparison of the inferred proper motions of the Quintuplet cluster with those from published work. (2) Show evidence for the first new co-moving group - and thus a potential cluster - in the nuclear bulge. We chose to focus our pilot study on field 19 of GALACTICNUCLEUS survey characterised by H{\scriptsize II} regions \citep{Dong-2017-hii}, that points towards the presence of young, hot stars.

In Sect. \ref{section:Observations} the data and the methodology of treating data are described. In
Sect. \ref{section:results} we present the results of the proper motion measurements of our sources and their statistical analysis. We also compare our proper motion results with those of the massive stars of the Quintuplet cluster. We summarise and discuss our results in Sect. \ref{section:summary}.


\section{Observations and methodology}
\label{section:Observations}

We use imaging data of the GALACTICNUCLEUS survey (GNS) \citep{Nogueras2018prep} from the High Acuity Wide-field K-band Imager (HAWK-I) at the Very Large Telescope (VLT) with the Paschen-$\alpha$ survey (P$\alpha$S) \citep{Wang:2010fk, Dong:2011ff} data from the Near-Infrared Camera and MultiObject Spectrometer (NICMOS) Camera~3 (NIC3) on the \textit{Hubble Space Telescope} (HST) in order to obtain the stellar proper motions. The time baseline between the two data sets is seven years (HAWK-I/VLT: epoch 2015, NICMOS/HST: epoch 2008).

\subsection{The GNS data}
\label{section:The GNS }

GNS is a $0.2''$ angular resolution survey of the GC in the NIR ($J$, $H$, and $K_{\rm s}$). Its high spatial resolution is due to using the holographic imaging technique \citep{Schodel:2013fk}. 
HAWK-I imager with the pixel scale of $0.106''$/pixel has four 2048 $\times$ 2048 pixel detectors (four chips).
Since we observed with short integration time exposures, the detector had to be windowed resulting in a field of view (FOV) of 2048 $\times$ 768 pixel for each chip. The distortion of the detector has been corrected by using the stellar positions from VVV survey and obtaining a distortion solution for different bands and fields. \textit{StarFinder} software package \citep{Diolaiti2000} has been used for astrometry and photometry. The sampling of $0.106''$ per pixel is not sufficient to have the images with $0.2''$ angular resolution, therefore the final images are rebinned with a factor of 2. The detailed description of the survey, data reduction, distortion correction, and source detection processes can be found in \cite{Nogueras2018a, Nogueras2018prep}. The astrometric uncertainty of the stars was obtained from the comparison of three independent data subsets \citep[as explained for photometric measurements in the paper by][]{Nogueras2018a}. The position uncertainties of one of the epochs studied in this work for HAWK-I/VLT observations (chip \#1 of pointing 19; see Table~1) are presented in Fig.~\ref{fig:pos_uncer}. We only considered the stars with a relative astrometric position uncertainty of less than 2~mas, for our proper motion analysis.

\begin{figure*}[!t]
  \includegraphics[width=1.\textwidth]{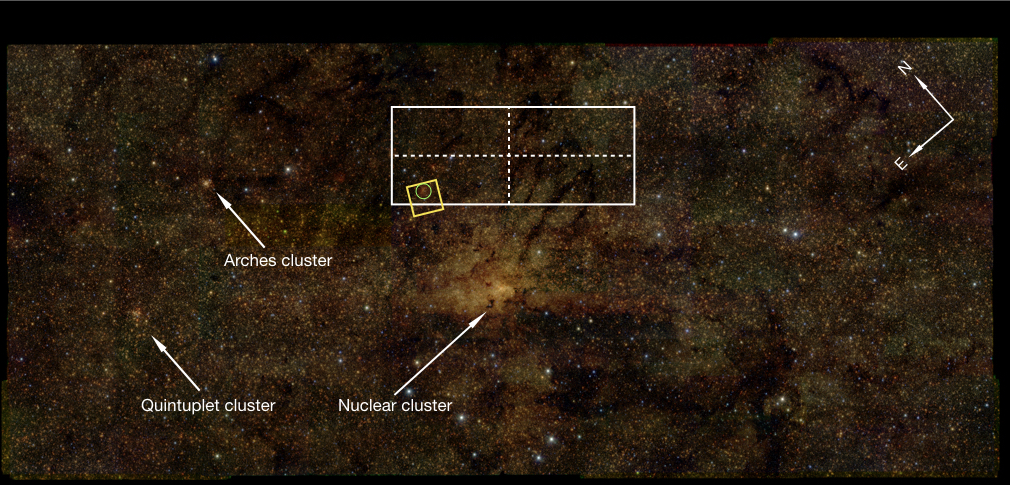}
  \caption{RGB image of the GC covering all the pointings of the GNS from \cite{Nogueras2018prep}. Blue stands for the $J$ band, green for the $H$ band, and red for the $K_{\rm s}$ band. The white rectangle shows pointing 19 of GNS with its 4 chips, with the size of $\sim 7.95' \times 3.43'$. The yellow square outlines one of the small fields of NICMOS/HST with the FOV of $51.2'' \times 51.2''$ and the green circle an H{\scriptsize II} region with strong Paschen-$\alpha$ emission labelled as H1 \citep{yusefzadeh1987}. The nuclear star cluster, the Quintuplet cluster, and the Arches cluster are also indicated.
}

  \label{fig:f19} 
\end{figure*}

In this paper, we use chip \#1 of pointing 19 (F19) and also chip \#2 of pointing 10 (F10). Each pointing has the size of $7.95' \times 3.43'$. F19 is located $\sim 6.36'$ north of Sgr~A* as shown in Fig.~\ref{fig:f19}. F10 is where the Quintuplet cluster is located. Table 1 presents the data sets used in this work. We have considered only the $H$-band data because stars brighter than $K_{\rm s} \sim 11$ are strongly saturated in $K_{\rm s}$ and the extreme extinction in $J$ ($A_{J} \gtrsim7$) means that only the brightest stars are detected in $J$ with good signal-to-noise-ratio \citep{Nogueras2018a, Nogueras2018prep}.

\begin{figure}[!htb]
  \includegraphics[width=0.45\textwidth]{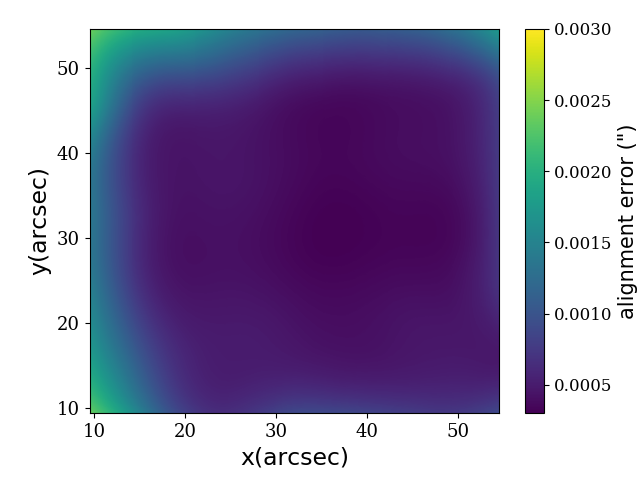}
  \caption{Interpolated alignment uncertainty map as a function of position in one of the small fields of NICMOS/HST shown in Fig.~1.}
  \label{fig:interpolation} 
\end{figure}


\begin{figure}[!htb]
  \includegraphics[width=\columnwidth]{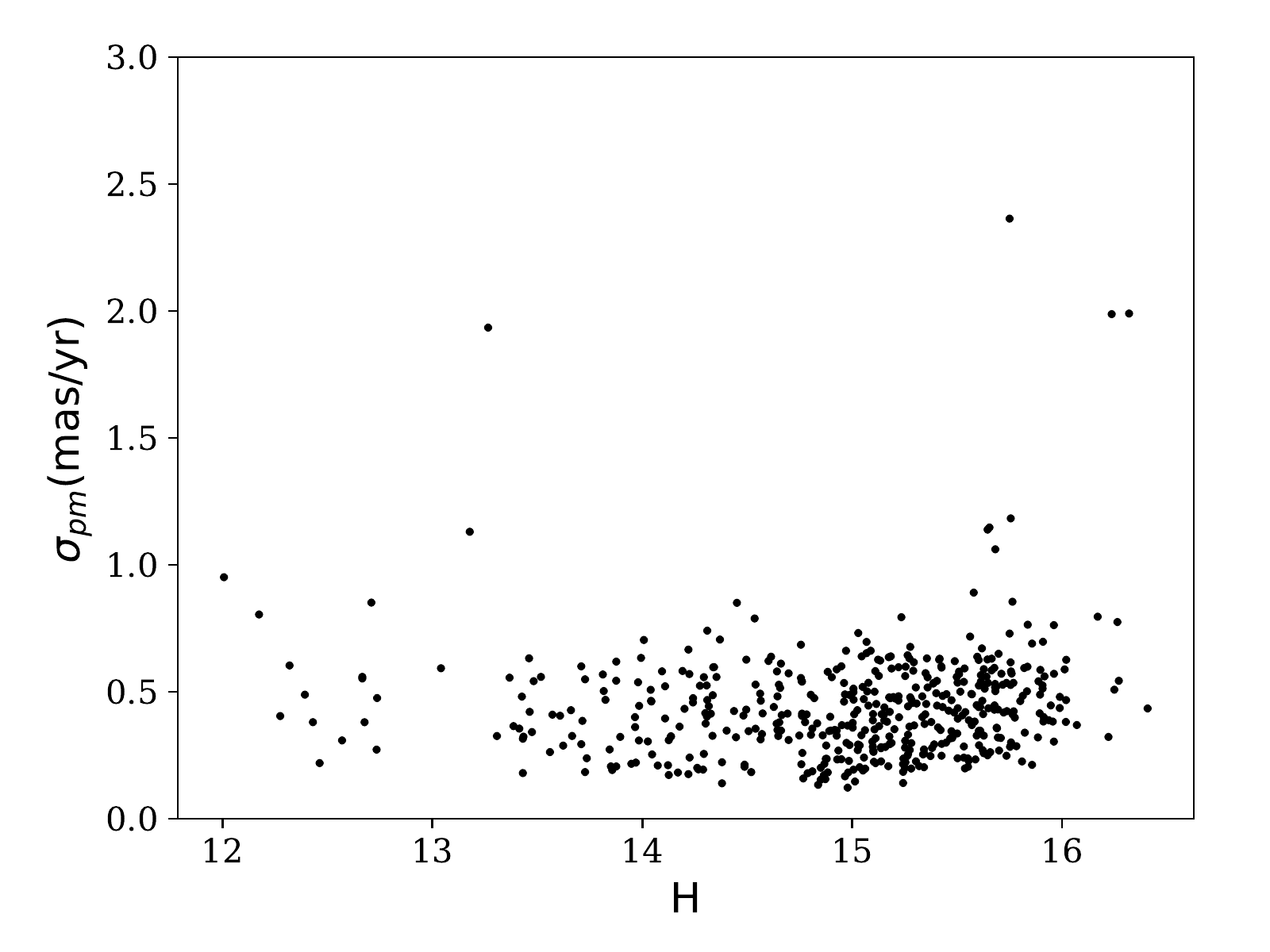}
  \caption{The uncertainties of stellar proper motions in chip \#1 of F19 as a function of H magnitude. The uncertainties include astrometric uncertainties of each epoch added quadratically to the alignment uncertainties.  }
  \label{fig:dpm-f19-chip1}
\end{figure}



    \begin{figure*}[]
      \begin{center}
    
        \subfigure{%
            \includegraphics[width=0.45\textwidth]{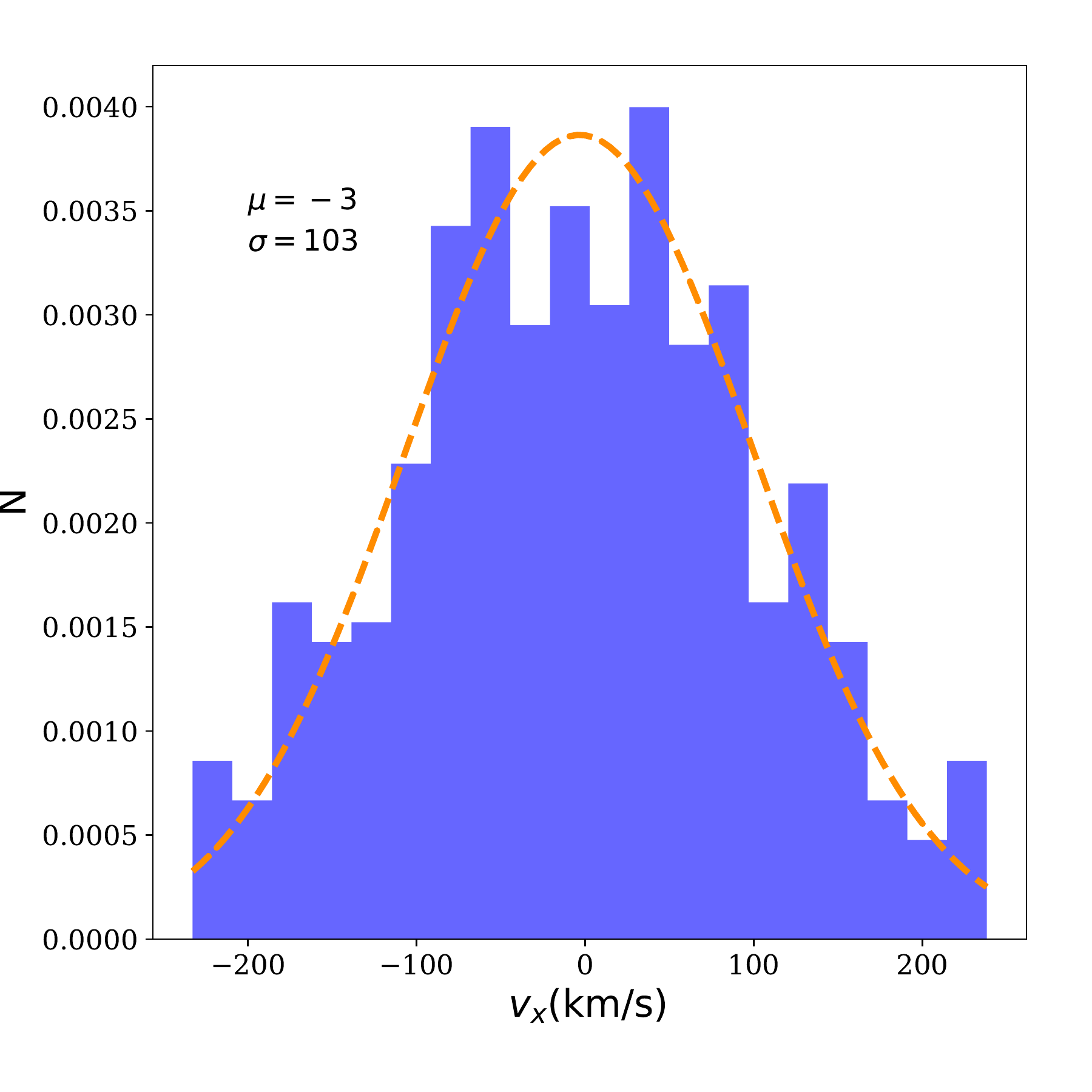}
        }
        \subfigure{%
          \includegraphics[width=0.45\textwidth]{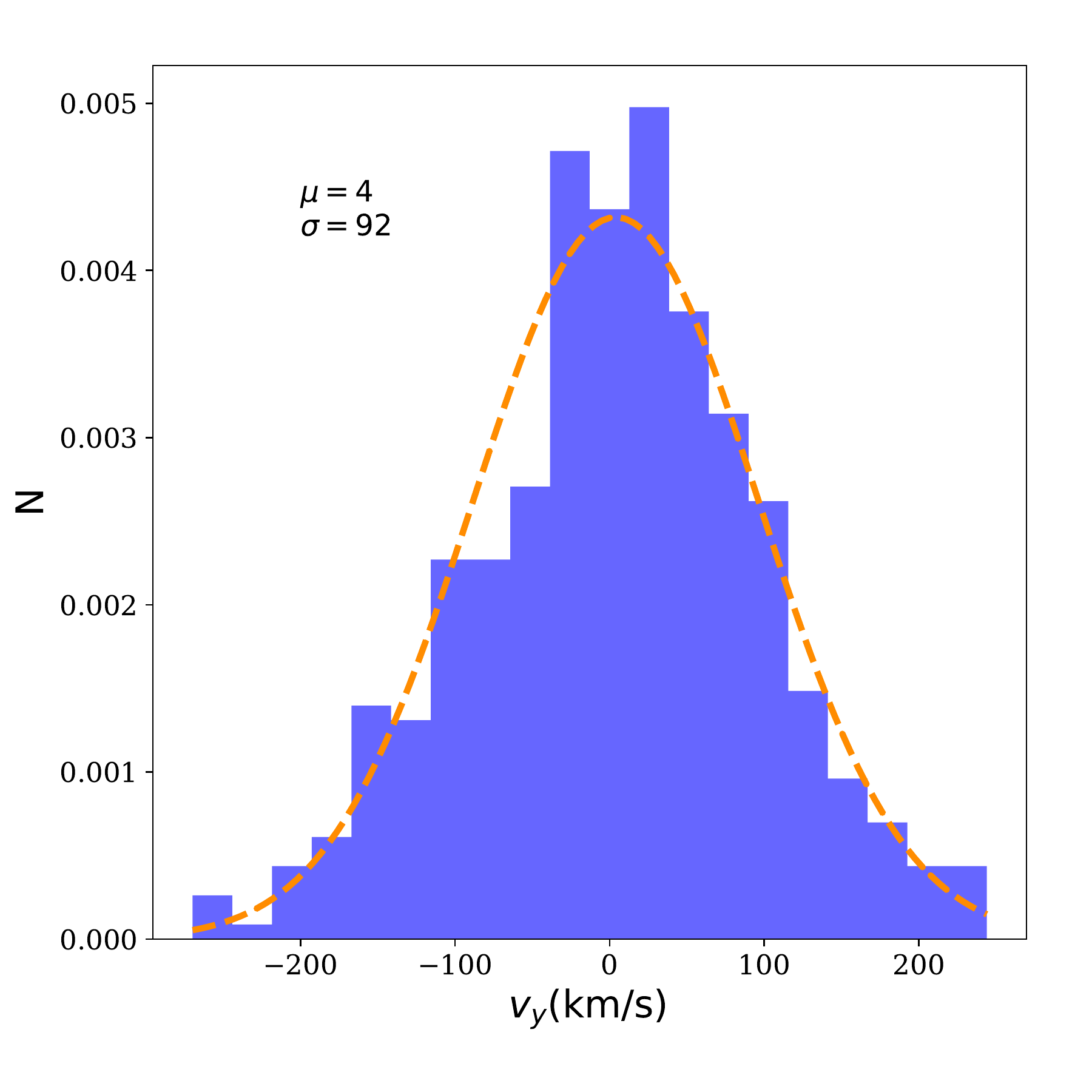}
        }\\

     \end{center}
    
    \caption{Normalised velocity distributions of stars in F19 chip \#1 in two directions: along (left panel) and perpendicular (right panel) to the Galactic Plane. The mean and standard deviation of the distributions are shown in $\mathrm{km\, s^{-1}}$ as $\mu$ and $\sigma$. The dashed orange lines are the fits to the histograms with Gaussian functions. }
\label{fig:hist_f19}
\end{figure*}


\subsection{The P$\alpha$S data}
\label{section:PalphaS survey data}

The P$\alpha$S maps the central $36' \times 15'$ of the GC with narrow-band filters $F187N$ and $F190N$. NIC3 has a FOV of $51.2'' \times 51.2''$. Since we are interested on measuring proper motions, we use the latter band that mostly includes the stellar continuum at 1.90 $\mu m$. The final mosaic has a pixel size of $0.1''$ with a spatial resolution of $0.2''$. The survey, the data analysis method, and the full source list catalogue is described in details in \cite{Dong:2011ff}. We extracted the point spread function (PSF) and detected the sources with \textit{StarFinder} for each position image, and obtained the astrometry uncertainties from this software. The relative astrometric uncertainties of one of the pointings studied here are presented in Fig.~\ref{fig:pos_uncer_hst}. For our proper motion analysis we only considered the detected stars with less than 3~mas position uncertainty.

As a comparison, GNS has a larger FOV per pointing
($\sim$ 37 times larger) and the same angular resolution $\sim0.2''$ as compared to the P$\alpha$S.

\begin{table}[b]
\caption[Observation details of the HAWK-I fields used in this work.]{Observation details of the HAWK-I fields used in this work.}
\begin{threeparttable}
\scriptsize
\centering
\begin{tabular}{c c c c c c c}
\hline
\hline 
Field & Observing date & Filter & seeing & N\tnote{a} & NDIT\tnote{b} & DIT\tnote{c} \\[0.5ex]
&(d/m/y)&&(arcsec)&&&(sec)\\
\hline
 &  27/06/2015 & \textit{J} & 0.34 & 49 & 20 & 1.26 \\
F19 &27/06/2015 & \textit{H} & 0.38 & 49 & 20 & 1.26\\
&  02/07/2015 & \textit{$K_{s}$} & 0.45 & 49 & 20 & 1.26 \\[0.2ex]
\hline 
&08/06/2015 & \textit{J} & 0.35 & 49 & 20 & 1.26 \\
F10&08/06/2015 & \textit{H} & 0.46 & 49 & 20 & 1.26\\
&08/06/2015 & \textit{$K_{s}$} & 0.45 & 49 & 20 & 1.26 \\[0.2ex]%

\hline
\end{tabular}

\begin{tablenotes}
\small
\item [a]Number of pointings.\item[b]Number of frames per pointing.\item[c]Integration time for each frame. The total integration time of each observation is given by $\textit{N} \times NDIT \times DIT$.

\end{tablenotes}

\end{threeparttable}
\label{table:data}
\end{table}

\subsection{Transformation procedure}
\label{section:Methodology}

Since we do not have any absolute astrometric reference for our fields, we assume that the stellar proper motions cancel out, on average. We have used this assumption to transform the stellar positions into a common reference frame.
Since the FOV of the NICMOS/HST detector is relatively small (see Fig.~\ref{fig:f19}), aligning the stellar positions to a common reference frame can introduce uncertainties. Therefore, we need to be cautious aligning the positions in order to measure accurate proper motions. Our alignment procedure is based on \cite{schoedel2009} which we summarise here.
We considered our reference frame to be the stellar positions from the HAWK-I/VLT 2015 epoch. From a primary common list of stars, we selected the transformation reference stars.

Then we sampled our reference stars uniformly to not be biased by the regions of higher stellar density and to reduce the systematic uncertainties. Therefore, we selected the reference stars of GNS on a grid of 36~$\times$~36~pixel ($1.9'' \times 1.9''$), and the star with the smallest position uncertainty was selected in each grid field.
Subsequently we applied a polynomial transform of degree 3 to align the stellar positions of the detected stars in each of the small fields of the 1.90~$\mu m$ HST/NICMOS 2008 epoch to the reference frame. Smaller degrees of the polynomial were also tested, which resulted in somewhat less accurate results and more systematic changes throughout the field. The parameters of the polynomial transformation between the stellar positions were determined through an iterative least squares algorithm (IDL \textit{POLYWARP} procedure) by comparing the positions of the transformed stars with the positions of the corresponding stars in the reference frame \citep[see equations A.1 and A.2 in][]{schoedel2009}. 

In order to obtain the uncertainties of the alignment, we used Jackknife method as described in the following. We removed one of the stars at a time from the sample and repeated the alignment procedure for 1000 times obtaining new values for the transformation. As a result, the medians and the standard deviations of the resulting distributions of the transferred positions of stars were considered as the best values and uncertainties. In this way, we could infer the uncertainty of alignment for each star. Using the Jackknife procedure also helped us to take into account the systematics due to the reference star selection. 

We created the alignment uncertainty maps, that show the alignment uncertainty as a function of position in the fields, by interpolation and considering the individual alignment uncertainties (see Fig.~\ref{fig:interpolation} for one of the fields as an example). The accuracy of the alignment depends on the field because of (a) variations in stellar surface density and (b) some of the NICMOS/HST pointings cover only partially the GNS pointings. The limiting factor in our analysis is the NICMOS/HST images because of their small FOV and low sensitivity.

\section{Results}
\label{section:results}

\subsection{Proper motions in F19 chip \#1}
\label{section:proper motions}

Eight images of NICMOS/HST have overlaps with the chip~\#1 of F19. After aligning the stars to a common reference frame, we produced the final stellar proper motions in the whole chip \#1 of F19 by merging all the proper motion measurements. Stars were matched between the epochs by checking around the position in the reference frame in a circle with $0.1''$ radius. Finally, after applying the methodology described above, proper motions of 481 stars were measured for this field by calculating the displacement of the positions of the stars in the overlapping region of the two epochs divided by their temporal baseline $\Delta$T = 7 years. To obtain uncertainties of the proper motion measurements, standard error propagation was applied by considering the individual position uncertainties in the directions of parallel and perpendicular to the Galactic Plane for each star in two epochs and adding them quadratically to the alignment uncertainties.

The proper motion outliers were removed by considering only the velocities $\lesssim2\times$weighted standard deviation of the individual star velocities from the weighted-average. In Fig.~3 the proper motion uncertainties of this field are shown after removing the outliers. In the following we refer to the proper motion measurements after removing the outliers as significant values. Figure~\ref{fig:hist_f19} shows the velocity distributions of stars in F19 chip \#1 with considering the significant values. The measured velocity dispersion of all sources with significant proper motions (446 stars) in the field along the Galactic Plane is $103~\mathrm{km\, s^{-1}}$ ($2.575~\mathrm{mas\, yr^{-1}}$) and perpendicular to it $92~\mathrm{km\, s^{-1}}$ ($2.3~\mathrm{mas\, yr^{-1}}$). The measured velocity dispersion is higher than the intrinsic one because it is folded with the uncertainty of the proper motion measurements.


    \begin{figure*}[!htb]
      \begin{center}
    
        \subfigure{%
            \includegraphics[width=0.45\textwidth]{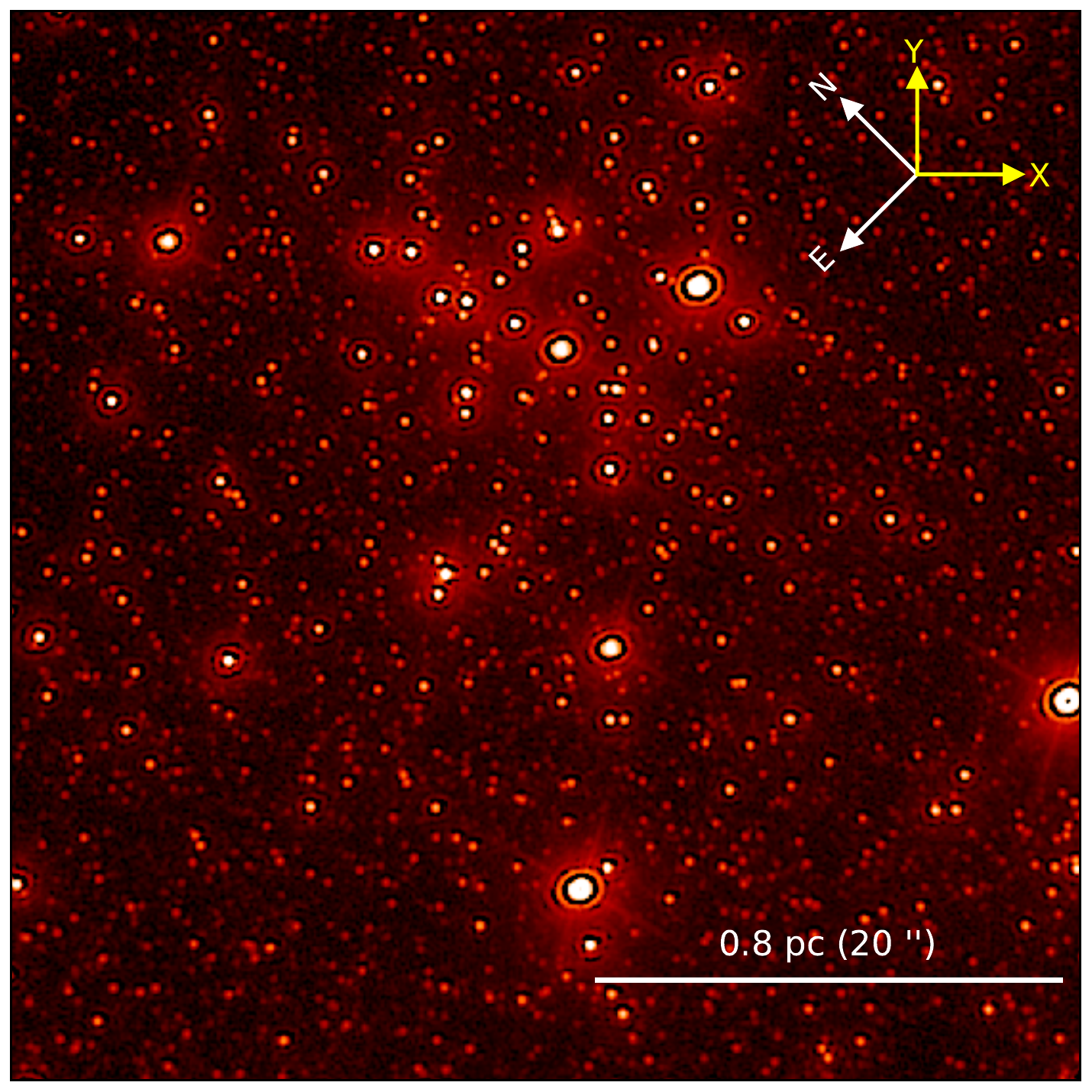}
        }
        \subfigure{%
          \includegraphics[width=0.45\textwidth]{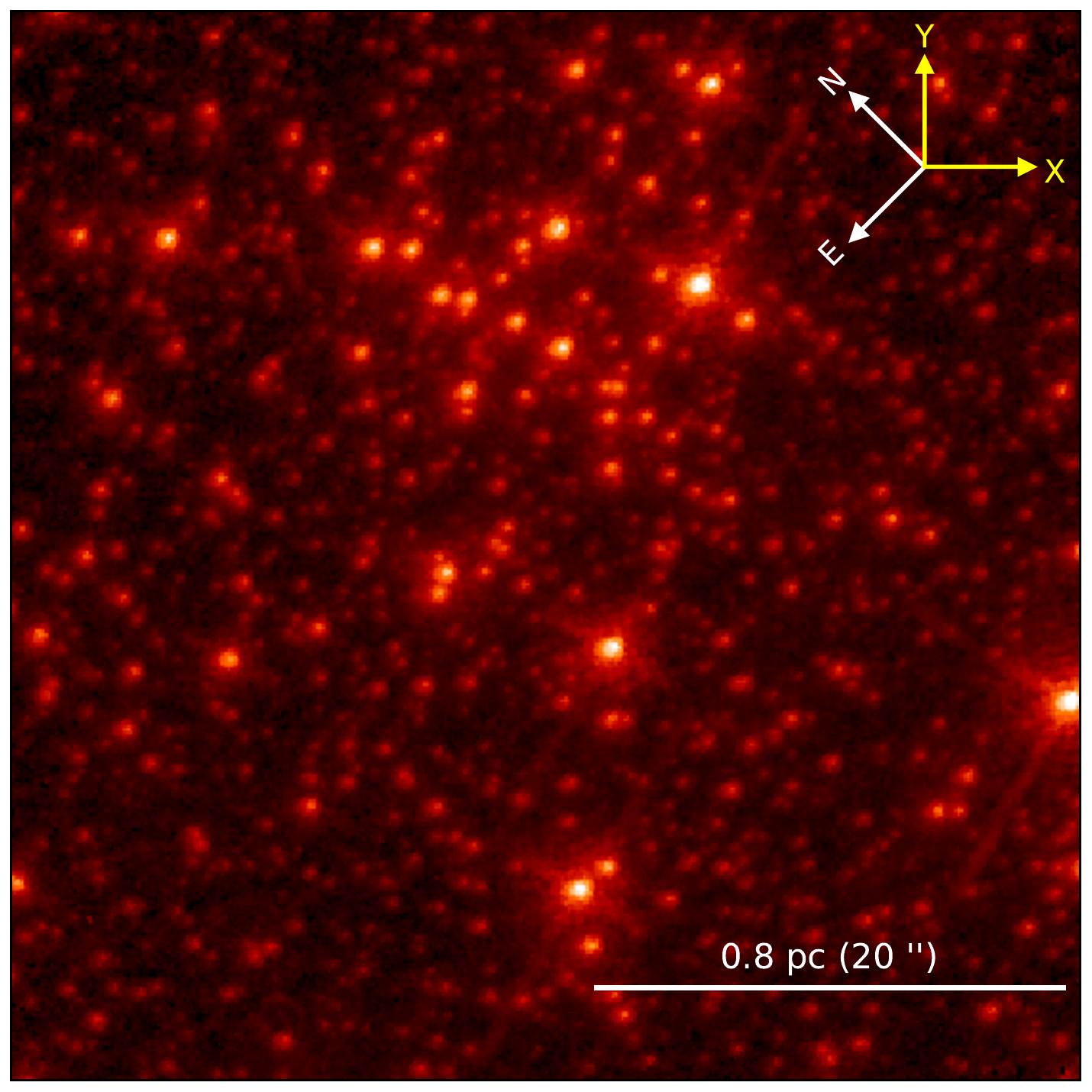}
        }\\

     \end{center}
    \caption{Quintuplet cluster from HAWK-I/VLT (left) and NICMOS/HST (right). We calculate the stellar proper motions of the presented region.}
\label{fig:hawki-hst}
\end{figure*}


\begin{figure}[]
  \includegraphics[width=\columnwidth]{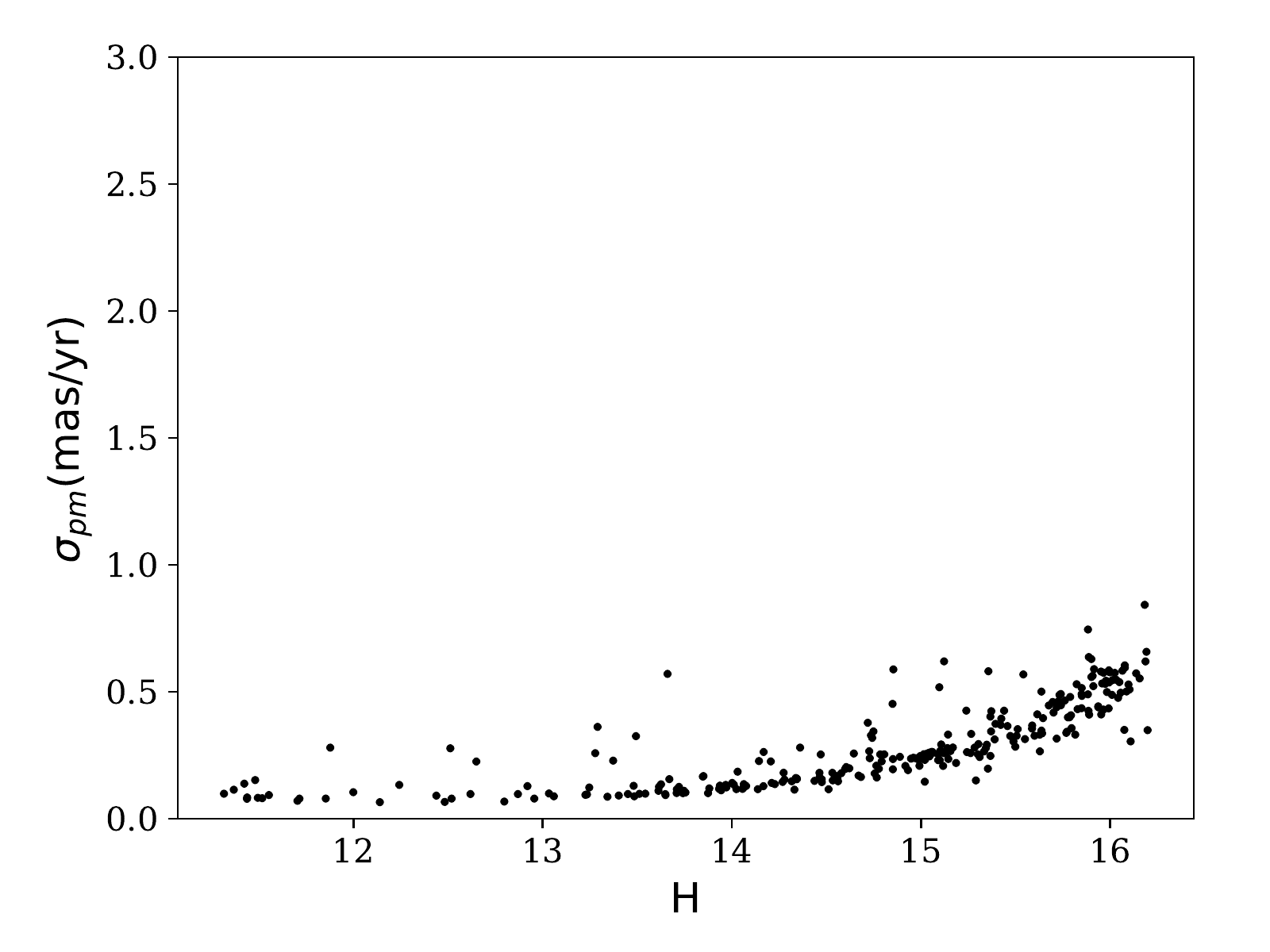}
  \caption{The uncertainties of stellar proper motions of the Quintuplet cluster as a function of H magnitude. The uncertainties include astrometric uncertainties of each epoch added quadratically to the alignment uncertainties.}
  \label{fig:dpm-quin}
\end{figure}



 \begin{figure}[]
  \includegraphics[width=\columnwidth]{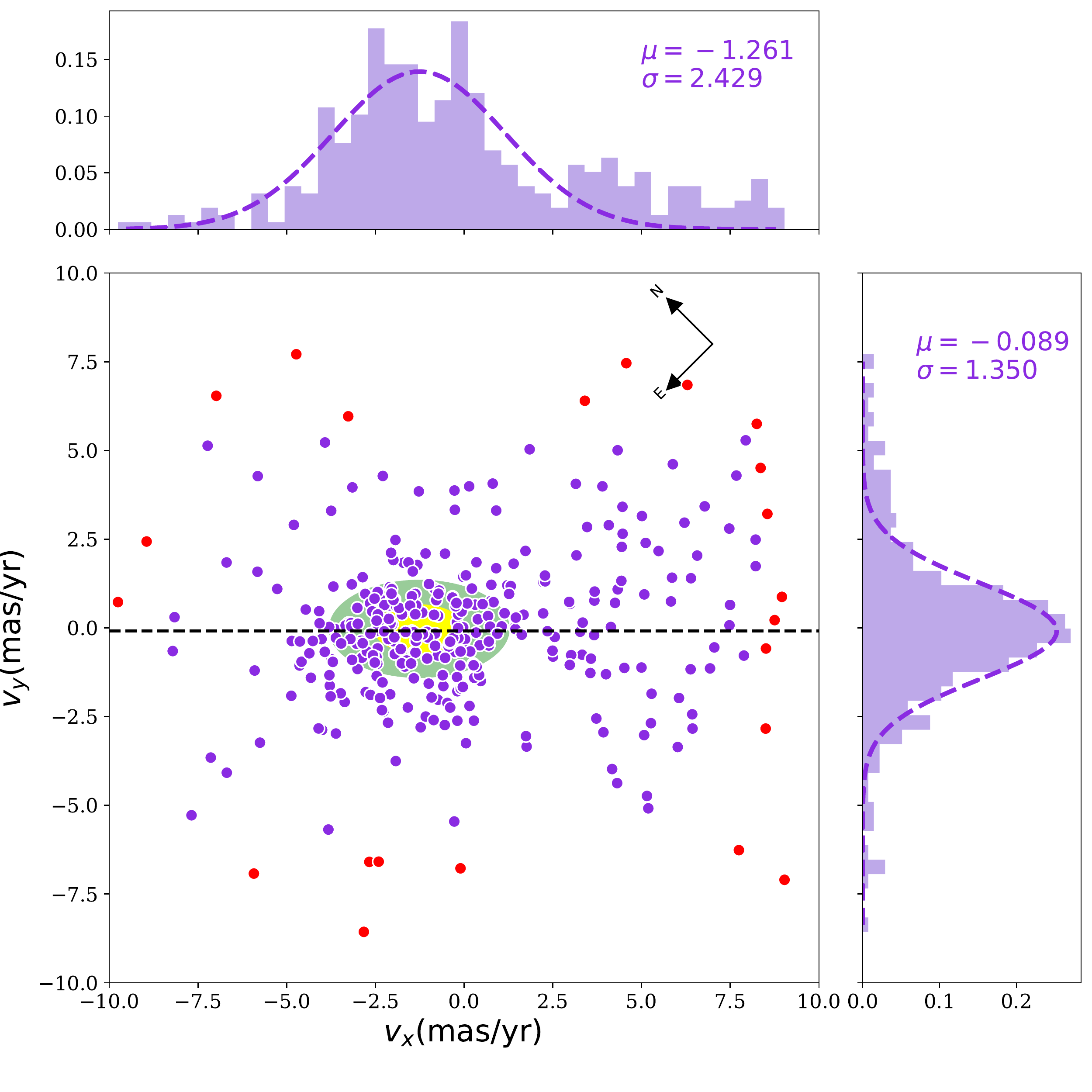}
  \caption{Vector-point diagram of the proper motions for the region of the Quintuplet cluster. The yellow and green ellipses show 1$\sigma$ and 2$\sigma$ distributions for a 2D Gaussian fit respectively. The dashed line shows the Galactic plane orientation. Top and right panels present the histogram distributions of $v_x$ and $v_y$ , as well as their Gaussian fits with $\mu$ and $\sigma$ showing the mean and standard deviation of the distributions in $\mathrm{mas\, yr^{-1}}$. The motion parallel to the Galactic plane ($v_x$) presents a tail of field sources, whereas the motion perpendicular to the plane ($v_y$) is dominated by velocity dispersion. Violet points show the stars with significant proper motions and the red points the non-significant ones.
  }
  \label{fig:quintuplet_contour} 
\end{figure}



 \begin{figure}[]
  \includegraphics[width=\columnwidth]{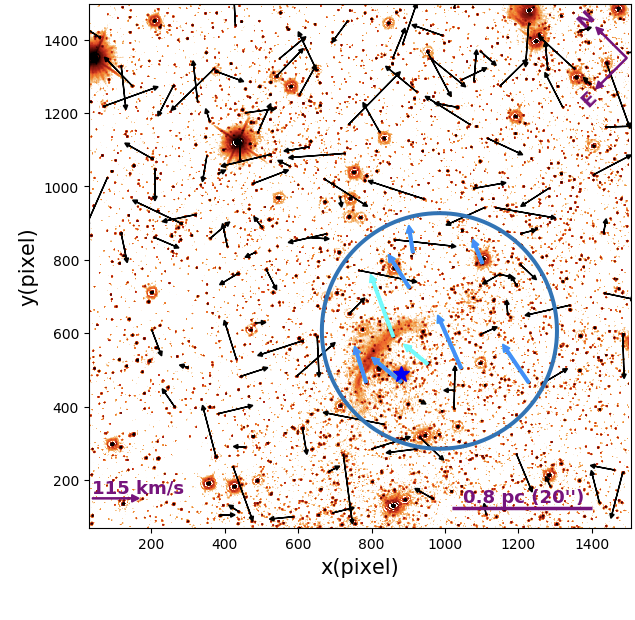}
  \caption{Proper motion measurements from GNS and P$\alpha$S. The blue arrows surrounded by a blue circle present the co-moving group of stars in the H1 H{\scriptsize II} region. The cyan arrows show the foreground stars identified with the CMD, which we do not include in our cluster analysis. The blue star shows a Paschen $\alpha$ emitting source \citep{Dong-2017-hii}.} 
  \label{fig:pm_arrow_overplot} 
\end{figure}


    \begin{figure}[!t]

            \includegraphics[width=\columnwidth]{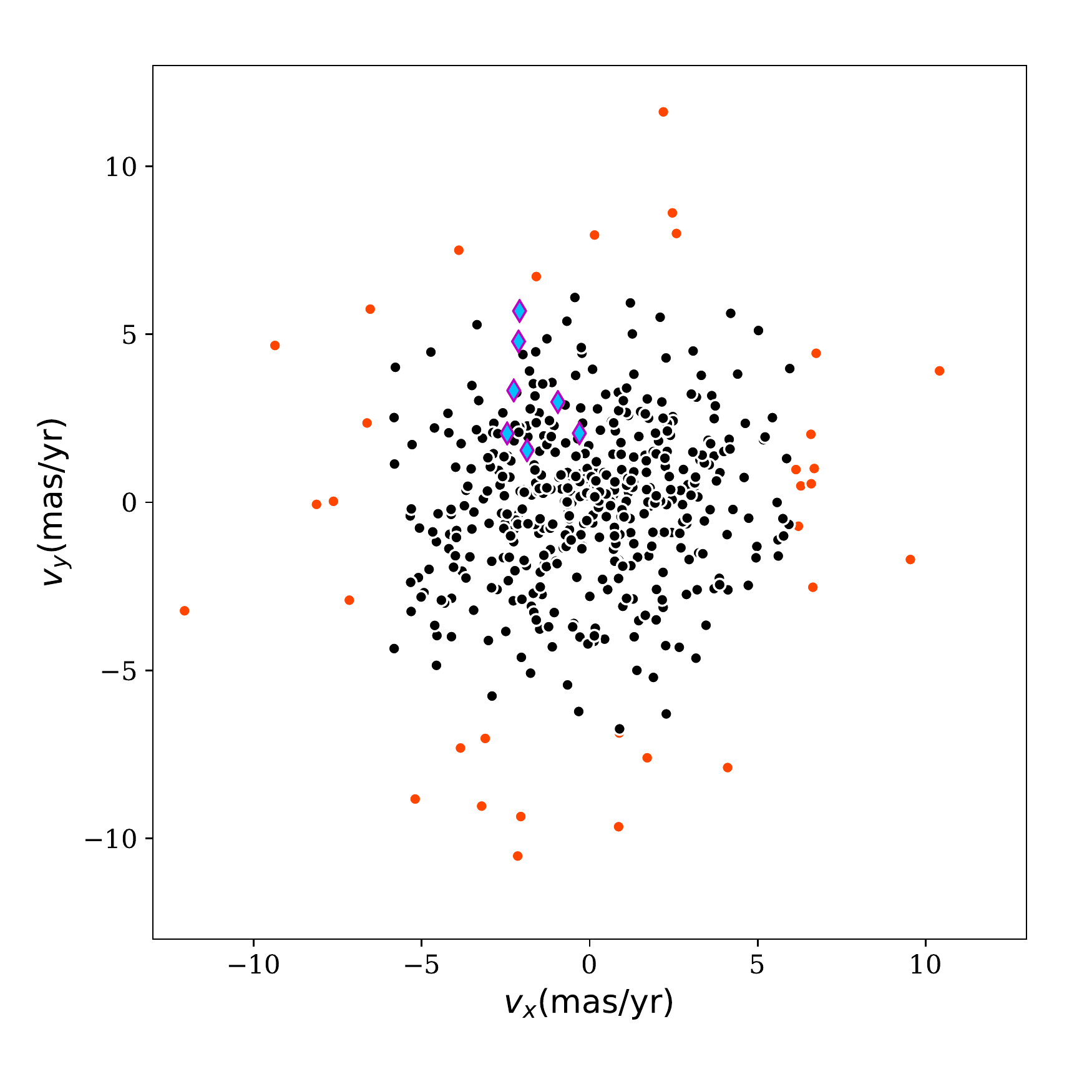}

    \caption{Vector-point diagram of the stars in chip \#1 of F19. Total of 481 stars, with black points showing the stars with significant proper motions and the red points the non-significant ones. The proper-motion dispersion in x axis is $2.58~\mathrm{mas\, yr^{-1}}$ and in y axis is $2.3~\mathrm{mas\, yr^{-1}}$. The blue diamonds show the group of co-moving stars.}
\label{fig:vector-point}
\end{figure}


 \begin{figure}[]
  \includegraphics[width=\columnwidth]{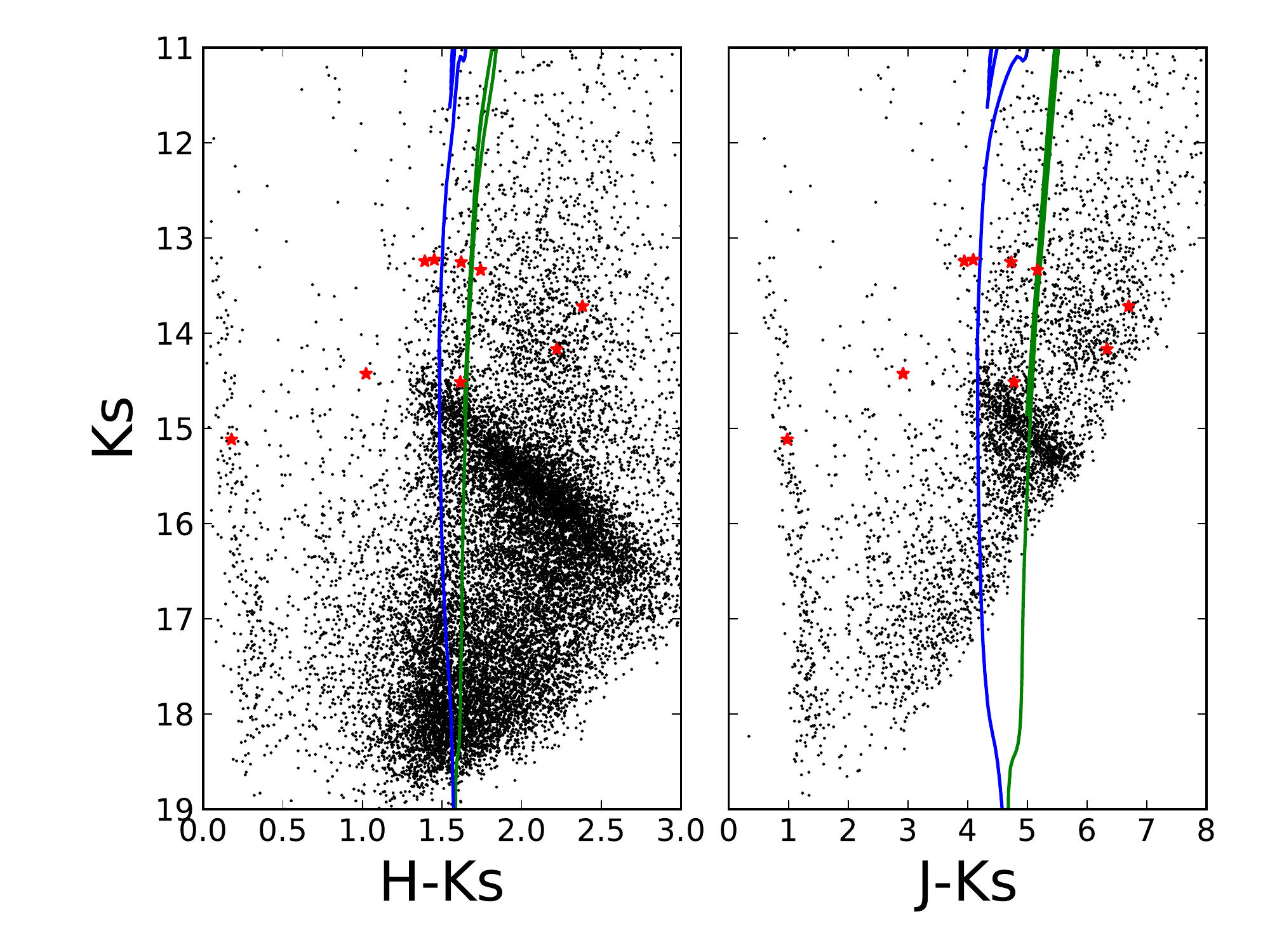}
  \caption{The CMD of the stars in chip \#1 of F19 with available HAWK-I/VLT magnitudes. The region with the highest density of stars indicates the red clump. The co-moving group of stars are marked in red. Two BaSTI isochrones of 30 Myr (blue) and 8 Gyr (green) are presented in each plot (see text in Sect.~\ref{section:co-moving stars}).
 }
 \label{fig:CMD_hawki}
\end{figure}


\subsection{Method validation}
\label{section:Method validation}

Proper motion studies have been used to identify cluster members in the Quintuplet cluster \citep{Stolte:2008uq, Stolte:2014ys, hosek:2015, rui:2019}. In order to test if our proper motion measurements are robust, we measured the proper motions of the Quintuplet cluster by using the HST data in the field that covers this cluster and the corresponding field of GNS (chip \#2 of F10; see Fig.~\ref{fig:hawki-hst}) following the same methodology described in Sects. \ref{section:Methodology} and \ref{section:proper motions}. In Fig.~\ref{fig:dpm-quin} the measured proper motion uncertainties of the Quintuplet cluster are shown. As expected, an increase in the uncertainty towards fainter magnitudes can be noticed. The difference between proper motion uncertainties of the Quintuplet cluster (Fig.~\ref{fig:dpm-quin}) with the ones of the F19 chip~\#1 (Fig.~3), is due to the Quintuplet pointing being a field with a very high density of relatively bright reference stars.

The obtained proper motion diagram is presented in Fig.~\ref{fig:quintuplet_contour} with a cluster population seen close to the origin and the field star population extended along the direction of the Galactic plane shown with a dashed line (see Fig. 7 of \cite{stolte:2015} for a comparison). The shift of the cluster population along the Galactic Plane direction from the origin is expected since the vector-point diagram is not centred on the cluster reference frame. The histograms of motion parallel and perpendicular to the Galactic Plane (x and y) are also shown in Fig.~\ref{fig:quintuplet_contour}. The peak of the cluster population in motion along the Galactic plane together with the extended tail of the field stars is visibly clear. The motion parallel to the Galactic Plane ($v_x$) includes a tail of field sources, while the motion perpendicular to the plane ($v_y$) is dominated by velocity dispersion. Our proper motion result is consistent with the previous studies of this cluster \citep[e.g. see Figs.~7 and 9 in][]{stolte:2015}, showing the feasibility of our approach. \\


\subsection{Co-moving stars}
\label{section:co-moving stars}

Figure~\ref{fig:pm_arrow_overplot} presents the significant proper motions of stars in part of chip \#1 of F19. We have detected, by eye, a group of co-moving stars that we mark on the HAWK-I image (see also Fig.~\ref{fig:pm_arrow_comove} which includes uncertainties as well). There is an H{\scriptsize II} region labelled as H1 \citep{yusefzadeh1987,Zhao93, Dong-2017-hii} with a Paschen $\alpha$ emitting source located in this field which must be associated with a massive star that has left the main sequence. Lyman continuum radiation from massive stars produces the Paschen $\alpha$ emission that traces the warm ionized gas. This gas could be from the stars, their immediate environment and surrounding ISM.

The co-moving stars that we identified by eye are also presented in the upper left quadrant of the vector-point diagram (Fig.~\ref{fig:vector-point}). We cannot measure the intrinsic velocity dispersion for this group of stars since it is dominated by measurement uncertainties. 
These stars are not easily identifiable as a clump in this diagram, contradictory to the similar diagram for the Quintuplet cluster, because of their small number. However, they occupy a similarly narrow space in the diagram than the Quintuplet stars.

In order to check if some of the stars in the co-moving group of stars are foreground stars and to exclude them from the subsequent analysis, we made CMDs. Figure~\ref{fig:CMD_hawki} presents the CMDs $K_{\rm s}$ vs. $H\mbox{-}K_{\rm s}$ and $J\mbox{-}K_{\rm s}$ vs. $J$ of stars of chip \#1 of F19 from the GNS. The co-moving group of stars are shown in red in this figure, from which two sources with $H\mbox{-}K_{\rm s}$ < 1.3 are foreground stars \citep[see][]{Nogueras2018a}. The stars with $1.3 < H\mbox{-}K_{\rm s} < 2$ are consistent with being main sequence (MS) or post MS stars with ages of a few to a few tens of Myr and the ones at $H\mbox{-}K_{\rm s}$ > 2 may be intrinsically reddened cluster members. We used BaSTI\footnote{http://basti.oa-teramo.inaf.it} non $\alpha$-enhanced models \citep{Pietrinferni:2004, Pietrinferni:2006} with solar metallicity to simulate stellar population of our sources. There is significant bias and degeneracy in estimating the age of a stellar population from such a small number of stars in such a complex field, with high extinction that can vary on arcsecond scales. The isochrones that we plot over the CMDs are therefore merely orientative. We show an 8 Gyr old isochrone for our sources along with a 30 Myr isochrone.

In the following, we continued our analysis without the foreground stars.

\subsection{Cluster analysis simulations}
\label{section:Cluster analysis simulations}

To verify the potential cluster nature of the co-moving group of stars, we need to identify that its stars are reasonably close in both velocity and position space. In order to identify that these stars move in space coherently and to obtain the significance of having a real group of stars, we used a Monte Carlo (MC) approach. 

 First, we defined a parameter \textit{velocity compactness ($dis\underline{~}v$)}, which shows how compact the distribution of stars in velocity space is, to be the sum of all possible pairwise distances of stars from each other in the Vector-point diagram in a given field (see Fig.~\ref{fig:vector-point}):

\noindent
\begin{equation}
dis\underline{~}v = \sum_{i=1}^{n} \sum_{j=1}^{n} \sqrt{(v_{xi} - v_{xj})^{2} + (v_{yi} - v_{yj})^{2}},
\end{equation}
where n is the number of stars. We calculated this value for the velocities of the co-moving group of stars. Then, we performed a MC simulation and calculated 1000 times \textit{$dis\underline{~}v$} for a simulated population of stars. For the simulations we assumed that the stars have the same positions as the observed ones. We assigned their velocities randomly by drawing them from normal distribution priors with initial parameters to be the median and $\sigma$ of the observed velocity distribution in both x and y directions. For each MC run, we searched for the most compact group of n stars in velocity space and calculated \textit{$dis\underline{~}v$} for it. We assumed n = the number of eye-selected co-moving stars.

 Moreover, we put a criterion for stars being close in position space, defined another parameter \textit{position closeness ($dis\underline{~}p$)} to be the sum of all possible pairwise distances of stars from each other in position space:
 
\noindent
\begin{equation}
dis\underline{~}p = \sum_{i=1}^{n} \sum_{j=1}^{n} \sqrt{(x_{i} - x_{j})^{2} + (y_{i} - y_{j})^{2}},
\end{equation}
where n is the number of stars. We calculated this value for the observed co-moving group of stars. We followed a MC approach this time for the positions and obtained 1000 times \textit{$dis\underline{~}p$} for a simulated group of stars.
 We assigned the stars positions randomly by drawing them from uniform priors with lower and upper limits of the observed x and y positions. In order to calculate \textit{$dis\underline{~}p$}, we searched for the most compact group of n stars in position space in each MC run. We assumed n to be similar to the definition mentioned above.
 
\begin{figure}[]
  \includegraphics[width=\columnwidth]{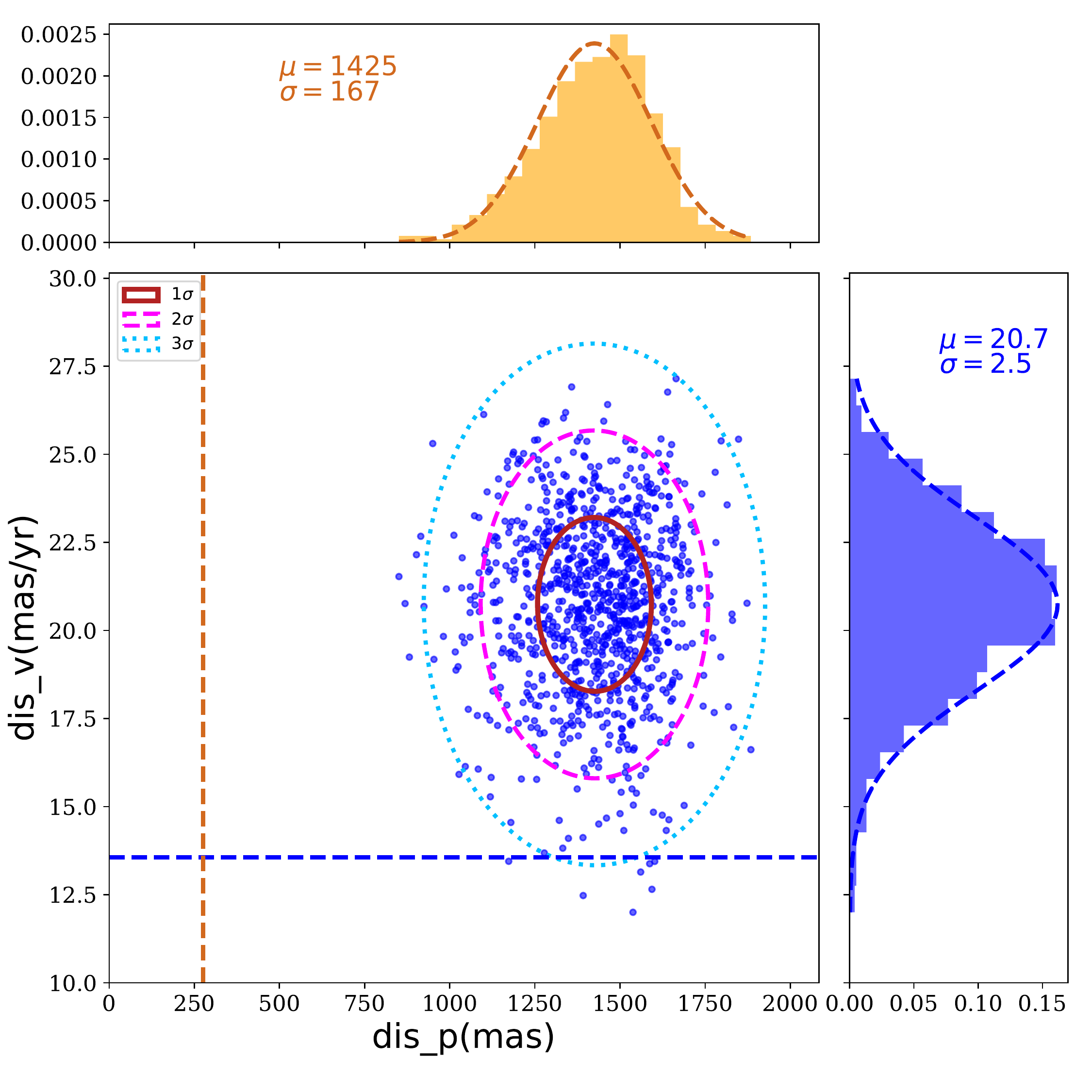}
  \caption{The distribution of the position closeness ($dis\underline{~}p$) and velocity compactness ($dis\underline{~}v$) for the simulated compact population of stars with its confidence intervals (1$\sigma$, 2$\sigma$, 3$\sigma$) together with the histograms and their Gaussian fits. The mean and standard deviation of the $dis\underline{~}p$ and $dis\underline{~}v$ distributions are shown as $\mu$ and $\sigma$, in mas and $\mathrm{mas\, yr^{-1}}$ respectively. The dashed orange and blue lines show respectively the observed $dis\underline{~}p$ and $dis\underline{~}v$ for the co-moving group of stars.}
 \label{fig:dis_marginal}
\end{figure}

Finally, we investigated the probability of finding the randomly closest group of stars to have the same velocity compactness and position closeness as the real data. As a result, Fig.~\ref{fig:dis_marginal} shows the distributions of \textit{$dis\underline{~}v$} and \textit{$dis\underline{~}p$} for the compact group of simulated stars together with the observed values of them for the discovered co-moving group presented by the dashed lines. 
 
 We find that \textit{$dis\underline{~}v$} and \textit{$dis\underline{~}p$} for the observed co-moving stars have more than $2\sigma$ and $\sim7\sigma$ offsets respectively from the mean value of their random distributions, therefore the co-moving group of stars is real. \\

\begin{figure}[]
  \includegraphics[width=\columnwidth]{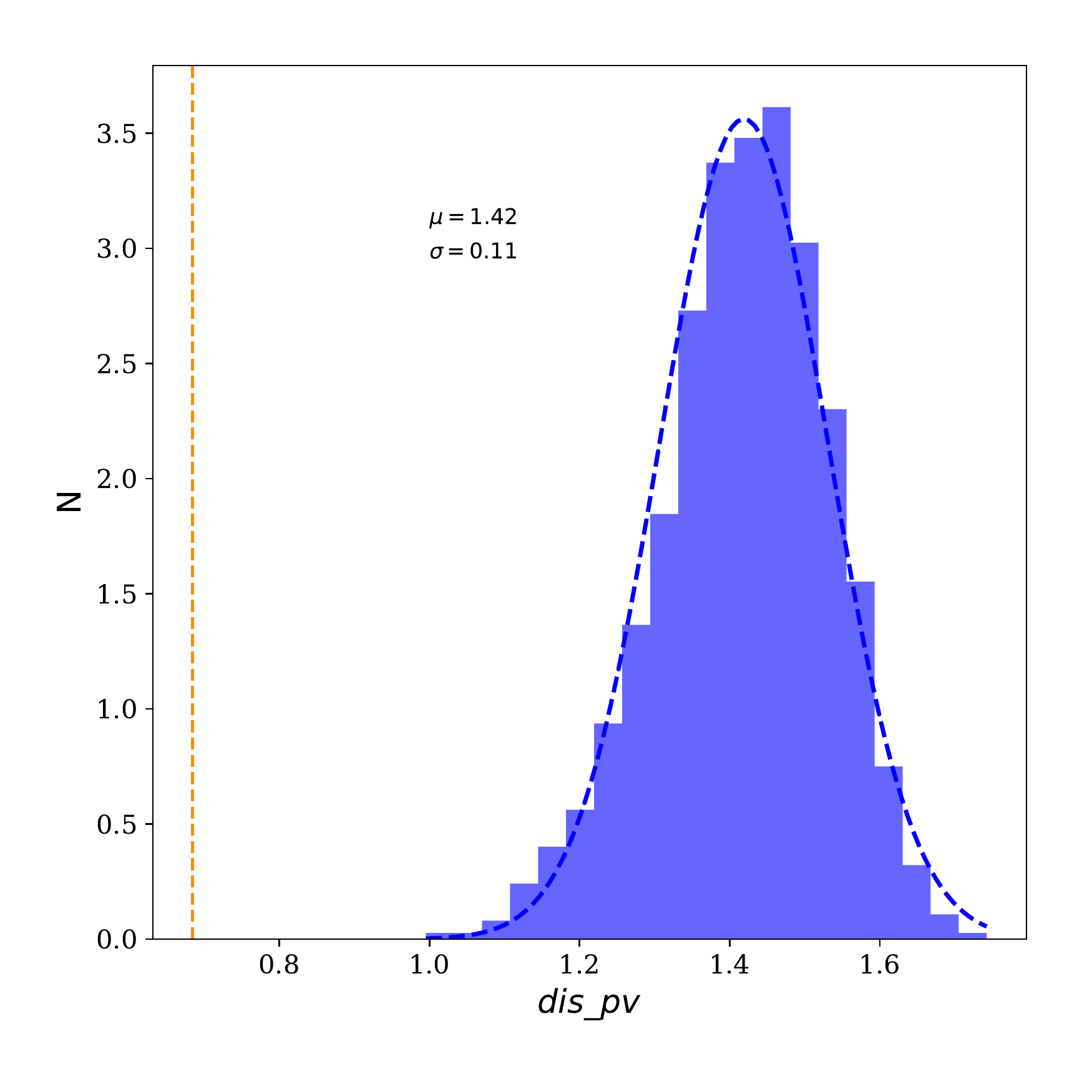}
  \caption{The distribution of $dis\underline{~}pv$ for the simulated compact population of stars. The dashed orange line shows the observed value for the co-moving group of stars.}
 \label{fig:hist_dist}
\end{figure}

 As an alternative method to what was described above we defined a parameter which includes normalised \textit{$dis\underline{~}v$} and \textit{$dis\underline{~}p$}:
 
 \noindent
 \begin{equation}
dis\underline{~}pv = \sqrt{(\frac{dis\_p}{\left< dis\_p \right>})^{2} + (\frac{dis\_v}{\left< dis\_v \right>})^{2}}, 
 \end{equation}
 to consider simultaneously both criteria of being close in velocity and position space for the group of stars. We calculated $dis\underline{~}pv$ for the co-moving group of stars and as shown in Fig.~\ref{fig:hist_dist}, it has more than $6\sigma$ offset from the mean value of the random distribution of $dis\underline{~}pv$ for the simulated compact population of stars. 
 We provide further tests to study the co-moving group of stars in Appendix B.



\section{Discussion and conclusions}
\label{section:summary}

Stellar proper motion measurements at the GC are required to address questions such as these: (1) What is the contribution of Bulge stars to the NSD? (2) What is the structure of the NSD? (3) There is solid evidence that the Initial Mass Function (IMF) is top-heavy at the GC \citep{Genzel:2010fk, hosek:2019}. Is the IMF top-heavy for all clusters formed recently at the GC? (4) Have the apparently isolated young massive stars throughout GC formed in isolation or are they accompanied by other young stars \citep{Dong:2011ff}?

The detection of young dissolved clusters in the GC is extremely difficult due to the high crowding and extinction.
Because of the extreme interstellar extinction in the region close to the Galactic Plane, the Gaia catalogue is incomplete and cannot detect the bulge and red giants but the brightest stars. Also, the GC NIR surveys such as VVV suffer from the source crowding in the GC \citep{schoedel2007} due to a limited seeing, saturation of the brightest sources ($K_{\rm s}\lesssim10$), and limited wavelength coverage.
However, GNS enables us to achieve high angular resolution and sensitivity through the holography technique, although saturation of brightest stars and degeneracy of CMDs due to the reddening still play roles. 

Therefore, in order to identify partially dissolved star cluster candidates in the nuclear disk we have initiated a study of stellar proper motions of this region by using the GNS and P$\alpha$S to overcome the degeneracy produced by reddening. Considering the 7 years baseline between the two data sets as well as the observational and alignment uncertainties, we could reach our aim of measuring proper motions with an accuracy of $\geq~1~\mathrm{mas\, yr^{-1}}$ (equivalent to $40~\mathrm{km\, s^{-1}}$ at the GC distance of 8~kpc). 

We verified the feasibility of our method by comparing the obtained proper motions of the Quintuplet cluster with the ones from the previous publications. We identified the first group of co-moving stars associated with an H{\scriptsize II} region nearby labelled as H1 through our proper motion analysis, which can trace a dissolved young cluster in the GC. Upon analysis of radial velocity distributions, \cite{Dong-2017-hii} claim that the configuration of the ionized gas in the H1 H{\scriptsize II} region can be interpreted by a bow-shock model, so that strong stellar wind of a quickly moving massive star compresses the molecular cloud and makes a thin shell.

Based on our several tests of significance analysis, the co-moving group of stars is a real group of stars that move coherently in space and may be the proof of the predicted, but so far undiscovered dissolving star clusters.
Their metallicity, age, and mass cannot be derived by our data, and future spectroscopic follow-up will constrain precise parameters of this group of stars.

We will continue working on completing our pilot proper motion study using the P$\alpha$S and GNS data. As we have shown in this work, the measured proper motions can be precise enough to pinpoint the locations of unknown young clusters. They can then be studied with high angular resolution imaging and spectroscopy follow-up observations (e.g. with ERIS/VLT) to constrain the rate and conditions of recent star formation at the GC, in particular the IMF.

\begin{figure}[!t]
  \includegraphics[width=\columnwidth]{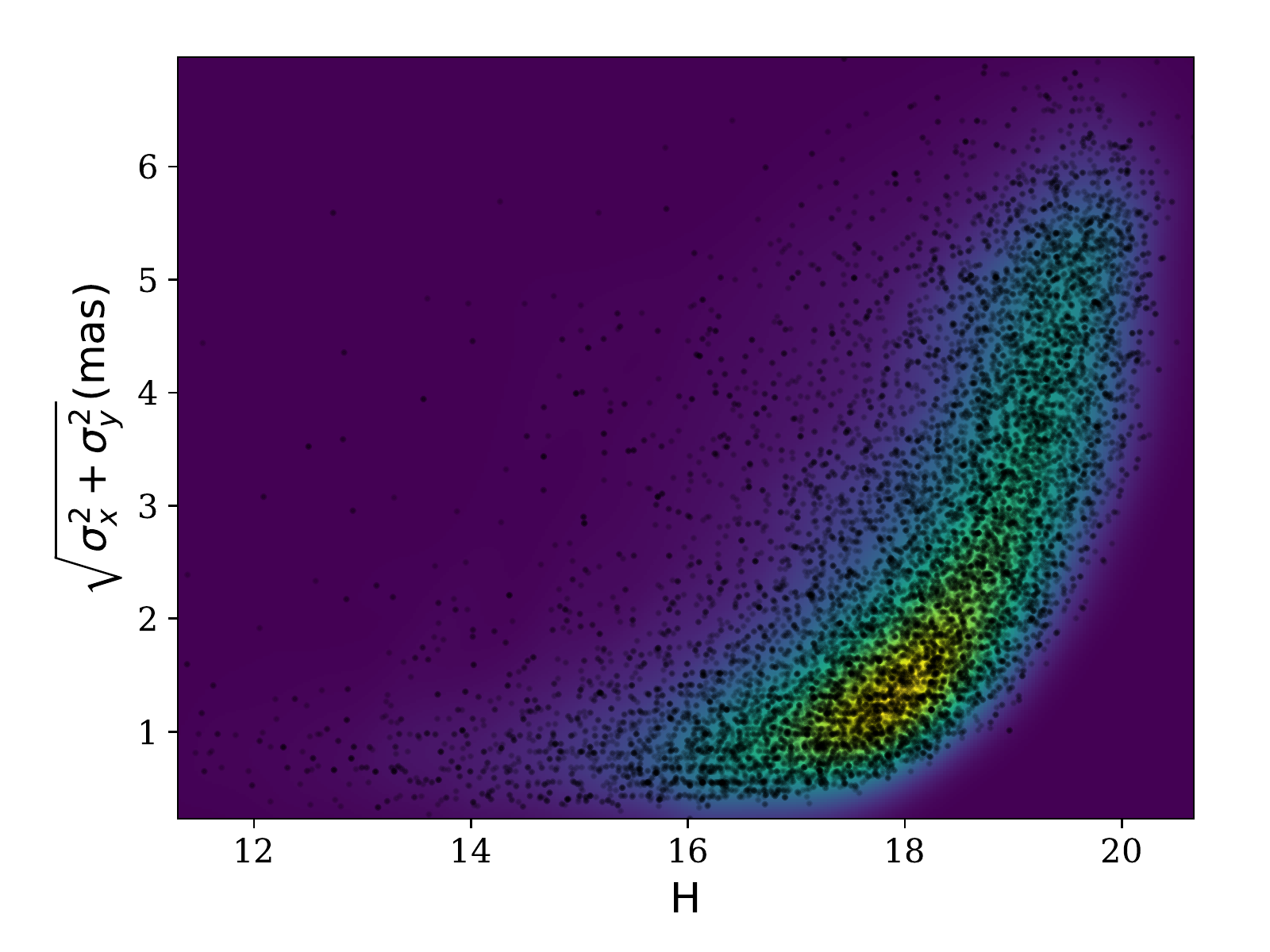}
  \caption{The density map of the relative astrometric uncertainties for a single chip and pointing of GNS. Showing only the uncertainties below $\sim$7 mas for illustration purposes.}
  \label{fig:pos_uncer_density} 
\end{figure}

Our pilot study is mainly limited by the small FOV of the NICMOS detector and the relatively low signal-to-noise ratio (SNR) of the P$\alpha$S narrow band images. Our next step will therefore consist of mapping the GNS region again, possibly repeatedly, with HAWK-I/VLT to measure the proper motions. The SNR of the GNS data is very high (on the order of a few 100 for $H\approx18$ stars). In Fig.~\ref{fig:pos_uncer_density} we show the relative astrometric uncertainty of GNS $H$-band astrometry for a single chip and pointing: It is smaller than 1\,mas for stars $H\lesssim18$. Within the FOV of HAWK-I we will be able to use thousands of stars in this brightness range to precisely ($\lesssim 1$\,mas) align the epochs via third or higher order polynomials and therefore to overcome the problem of small field of the NICMOS detector compared to GNS. As a result, we can obtain accurate proper motions ($\lesssim0.4$\,mas\,yr$^{-1}$) for $\sim$100 times more stars than in our pilot study. Data of this quality will allow us to address the questions mentioned above, to provide a catalogue for the community for detailed spectroscopic follow-up studies, and to create chemo-dynamical models of the GC \citep[e.g.][]{Portail:2017}.

NIR space-based astrometry missions, such as JASMINE \citep{Gouda:2018} are expected to give us information about the kinematics of stars to understand the properties and formation history of the GC \citep[see][]{Matsunaga2018}.
HAWK-I/VLT proper motions will make an ideal complement to these missions, such as JASMINE. On the one hand, small-JASMINE will be limited to the brightest percent (or less) of the stars at the GC, but provides ultra-high precision astrometry ($\sim 20\,\mu$as) of bright stars ($H\lesssim12.5$) over a very large field (a few square degrees) and thus links the GC to the Bulge/Bar. HAWK-I/VLT, on the other hand, will provide kinematic measurements for a hundreds or even thousand times more stars, but over a much smaller field.

Finally, linking proper motion measurements with spectroscopic measurements \citep[APOGEE, e.g.][]{Schoenrich:2015} will provide the kind of rich phase-space data set that is needed to understand the formation and evolution of the Milky Way's nucleus.

\begin{acknowledgements}

The authors would like to thank the anonymous referee for the helpful comments on this paper. The research leading to these results has received funding from the European Research Council under the European Union’s Seventh Framework Programme (FP7/2007-2013) / ERC grant agreement no [614922]. This work is based on observations made with ESO Telescopes at the La Silla Paranal Observatory under programmes IDs 195.B-0283 and 091.B-0418. We thank the staff of ESO for their great efforts and helpfulness. B.Sh, R.S, F.N.L, E.G.C, and A.T.G.C acknowledge financial support from the State Agency for Research of the Spanish MCIU through the "Center of Excellence Severo Ochoa" award for the Instituto de Astrofisica de Andalucia (SEV-2017- 0709). F.N.L acknowledges financial support from a predoctoral contract of the Spanish Ministerio de Educacion, Cultura y Deporte, code FPU14/01700.

\end{acknowledgements}

\vspace*{0.5cm}
\bibliographystyle{aa} 
\bibliography{shahzamanian_publication.bib} 

\begin{thebibliography}{53}
\expandafter\ifx\csname natexlab\endcsname\relax\def\natexlab#1{#1}\fi

\bibitem[{{Blum} {et~al.}(2003){Blum}, {Ram{\'\i}rez}, {Sellgren}, \&
  {Olsen}}]{Blum:2003}
{Blum}, R.~D., {Ram{\'\i}rez}, S.~V., {Sellgren}, K., \& {Olsen}, K. 2003,
  \apj, 597, 323

\bibitem[{{Boehle} {et~al.}(2016){Boehle}, {Ghez}, {Sch{\"o}del}, {Meyer},
  {Yelda}, {Albers}, {Martinez}, {Becklin}, {Do}, {Lu}, {Matthews}, {Morris},
  {Sitarski}, \& {Witzel}}]{Boehle:2016zr}
{Boehle}, A., {Ghez}, A.~M., {Sch{\"o}del}, R., {et~al.} 2016, \apj, 830, 17

\bibitem[{{Clark} {et~al.}(2018){Clark}, {Lohr}, {Patrick}, {Najarro}, {Dong},
  \& {Figer}}]{clark:2018}
{Clark}, J.~S., {Lohr}, M.~E., {Patrick}, L.~R., {et~al.} 2018, \aap, 618, A2

\bibitem[{{Clarkson} {et~al.}(2012){Clarkson}, {Ghez}, {Morris}, {Lu},
  {Stolte}, {McCrady}, {Do}, \& {Yelda}}]{Clarkson:2012fk}
{Clarkson}, W.~I., {Ghez}, A.~M., {Morris}, M.~R., {et~al.} 2012, \apj, 751,
  132

\bibitem[{{Diolaiti} {et~al.}(2000){Diolaiti}, {Bendinelli}, {Bonaccini},
  {Close}, {Currie}, \& {Parmeggiani}}]{Diolaiti2000}
{Diolaiti}, E., {Bendinelli}, O., {Bonaccini}, D., {et~al.} 2000, \aaps, 147,
  335

\bibitem[{{Dong} {et~al.}(2017){Dong}, {Lacy}, {Sch{\"o}del}, {Nogueras-Lara},
  {Gallego-Calvente}, {Mauerhan}, {Wang}, {Cotera}, \&
  {Gallego-Cano}}]{Dong-2017-hii}
{Dong}, H., {Lacy}, J.~H., {Sch{\"o}del}, R., {et~al.} 2017, \mnras, 470, 561

\bibitem[{{Dong} {et~al.}(2011){Dong}, {Wang}, {Cotera}, {Stolovy}, {Morris},
  {Mauerhan}, {Mills}, {Schneider}, {Calzetti}, \& {Lang}}]{Dong:2011ff}
{Dong}, H., {Wang}, Q.~D., {Cotera}, A., {et~al.} 2011, \mnras, 417, 114

\bibitem[{{Eckart} \& {Genzel}(1996)}]{Eckart&Genzel1996}
{Eckart}, A. \& {Genzel}, R. 1996, \nat, 383, 415

\bibitem[{{Eckart} \& {Genzel}(1997)}]{Eckart&Genzel1997}
{Eckart}, A. \& {Genzel}, R. 1997, \mnras, 284, 576

\bibitem[{{Figer} {et~al.}(1999){Figer}, {McLean}, \& {Morris}}]{Figer:1999uq}
{Figer}, D.~F., {McLean}, I.~S., \& {Morris}, M. 1999, \apj, 514, 202

\bibitem[{{Figer} {et~al.}(2002){Figer}, {Najarro}, {Gilmore}, {Morris}, {Kim},
  {Serabyn}, {McLean}, {Gilbert}, {Graham}, {Larkin}, {Levenson}, \&
  {Teplitz}}]{Figer:2002qf}
{Figer}, D.~F., {Najarro}, F., {Gilmore}, D., {et~al.} 2002, \apj, 581, 258

\bibitem[{{Figer} {et~al.}(2004){Figer}, {Rich}, {Kim}, {Morris}, \&
  {Serabyn}}]{Figer:2004fk}
{Figer}, D.~F., {Rich}, R.~M., {Kim}, S.~S., {Morris}, M., \& {Serabyn}, E.
  2004, \apj, 601, 319

\bibitem[{{Fritz} {et~al.}(2011){Fritz}, {Gillessen}, {Dodds-Eden}, {Lutz},
  {Genzel}, {Raab}, {Ott}, {Pfuhl}, {Eisenhauer}, \&
  {Yusef-Zadeh}}]{Fritz:2011fk}
{Fritz}, T.~K., {Gillessen}, S., {Dodds-Eden}, K., {et~al.} 2011, \apj, 737, 73

\bibitem[{{Gallego-Cano} {et~al.}(submitted){Gallego-Cano}, {Schoedel},
  {Nogueras-Lara}, \& {et al.}}]{Gallego:2019}
{Gallego-Cano}, E., {Schoedel}, R., {Nogueras-Lara}, F., \& {et al.} submitted,
  \aap

\bibitem[{{Genzel} {et~al.}(2010){Genzel}, {Eisenhauer}, \&
  {Gillessen}}]{Genzel:2010fk}
{Genzel}, R., {Eisenhauer}, F., \& {Gillessen}, S. 2010, Reviews of Modern
  Physics, 82, 3121

\bibitem[{{Genzel} {et~al.}(2003){Genzel}, {Sch{\"o}del}, {Ott}, {Eisenhauer},
  {Hofmann}, {Lehnert}, {Eckart}, {Alexander}, {Sternberg}, {Lenzen},
  {Cl{\'e}net}, {Lacombe}, {Rouan}, {Renzini}, \&
  {Tacconi-Garman}}]{Genzel:2003it}
{Genzel}, R., {Sch{\"o}del}, R., {Ott}, T., {et~al.} 2003, \apj, 594, 812

\bibitem[{{Ghez} {et~al.}(2008){Ghez}, {Salim}, {Weinberg}, {Lu}, {Do}, {Dunn},
  {Matthews}, {Morris}, {Yelda}, {Becklin}, {Kremenek}, {Milosavljevic}, \&
  {Naiman}}]{Ghez2008}
{Ghez}, A.~M., {Salim}, S., {Weinberg}, N.~N., {et~al.} 2008, \apj, 689, 1044

\bibitem[{{Gillessen} {et~al.}(2009){Gillessen}, {Eisenhauer}, {Trippe},
  {Alexander}, {Genzel}, {Martins}, \& {Ott}}]{Gillessen2009stars}
{Gillessen}, S., {Eisenhauer}, F., {Trippe}, S., {et~al.} 2009, \apj, 692, 1075

\bibitem[{{Gouda}(2018)}]{Gouda:2018}
{Gouda}, N. 2018, in IAU Symposium, Vol. 330, Astrometry and Astrophysics in
  the Gaia Sky, ed. A.~{Recio-Blanco}, P.~{de Laverny}, A.~G.~A. {Brown}, \&
  T.~{Prusti}, 90--91

\bibitem[{{Gravity Collaboration} {et~al.}(2018){Gravity Collaboration},
  {Abuter}, {Amorim}, {Anugu}, {Baub{\"o}ck}, {Benisty}, {Berger}, {Blind},
  {Bonnet}, \& {Brandner}}]{gravity2018}
{Gravity Collaboration}, {Abuter}, R., {Amorim}, A., {et~al.} 2018, \aap, 615,
  L15

\bibitem[{{Hosek} {et~al.}(2015){Hosek}, {Lu}, {Anderson}, {Ghez}, {Morris}, \&
  {Clarkson}}]{hosek:2015}
{Hosek}, Matthew~W., J., {Lu}, J.~R., {Anderson}, J., {et~al.} 2015, \apj, 813,
  27

\bibitem[{{Hosek} {et~al.}(2019){Hosek}, {Lu}, {Anderson}, {Najarro}, {Ghez},
  {Morris}, {Clarkson}, \& {Albers}}]{hosek:2019}
{Hosek}, Matthew~W., J., {Lu}, J.~R., {Anderson}, J., {et~al.} 2019, \apj, 870,
  44

\bibitem[{{Launhardt} {et~al.}(2002){Launhardt}, {Zylka}, \&
  {Mezger}}]{Launhardt:2002nx}
{Launhardt}, R., {Zylka}, R., \& {Mezger}, P.~G. 2002, \aap, 384, 112

\bibitem[{{Maness} {et~al.}(2007){Maness}, {Martins}, {Trippe}, {Genzel},
  {Graham}, {Sheehy}, {Salaris}, {Gillessen}, {Alexander}, {Paumard}, {Ott},
  {Abuter}, \& {Eisenhauer}}]{Maness:2007}
{Maness}, H., {Martins}, F., {Trippe}, S., {et~al.} 2007, \apj, 669, 1024

\bibitem[{{Martins} {et~al.}(2008){Martins}, {Hillier}, {Paumard},
  {Eisenhauer}, {Ott}, \& {Genzel}}]{martins2008}
{Martins}, F., {Hillier}, D.~J., {Paumard}, T., {et~al.} 2008, \aap, 478, 219

\bibitem[{{Matsunaga}(2018)}]{Matsunaga2018}
{Matsunaga}, N. 2018, in IAU Symposium, Vol. 334, Rediscovering Our Galaxy, ed.
  C.~{Chiappini}, I.~{Minchev}, E.~{Starkenburg}, \& M.~{Valentini}, 57--64

\bibitem[{{Matsunaga} {et~al.}(2011){Matsunaga}, {Kawadu}, {Nishiyama},
  {Nagayama}, {Kobayashi}, {Tamura}, {Bono}, {Feast}, \&
  {Nagata}}]{Matsunaga:2011uq}
{Matsunaga}, N., {Kawadu}, T., {Nishiyama}, S., {et~al.} 2011, \nat, 477, 188

\bibitem[{{Neumayer}(2017)}]{Neumayer:2017}
{Neumayer}, N. 2017, in IAU Symposium, Vol. 316, Formation, Evolution, and
  Survival of Massive Star Clusters, ed. C.~{Charbonnel} \& A.~{Nota}, 84--90

\bibitem[{{Nishiyama} {et~al.}(2008){Nishiyama}, {Nagata}, {Tamura}, {Kandori},
  {Hatano}, {Sato}, \& {Sugitani}}]{Nishiyama:2008qa}
{Nishiyama}, S., {Nagata}, T., {Tamura}, M., {et~al.} 2008, \apj, 680, 1174

\bibitem[{{Nogueras-Lara} {et~al.}(2018){Nogueras-Lara}, {Gallego-Calvente},
  {Dong}, {Gallego-Cano}, {Girard}, {Hilker}, {de Zeeuw}, {Feldmeier-Krause},
  {Nishiyama}, {Najarro}, {Neumayer}, \& {Sch{\"o}del}}]{Nogueras2018a}
{Nogueras-Lara}, F., {Gallego-Calvente}, A.~T., {Dong}, H., {et~al.} 2018,
  \aap, 610, A83

\bibitem[{{Nogueras-Lara} {et~al.}(2019{\natexlab{a}}){Nogueras-Lara},
  {Sch{\"o}del}, {Gallego-Calvente}, {Dong}, {Gallego-Cano}, {Shahzamanian},
  {Girard}, {Nishiyama}, {Najarro}, \& {Neumayer}}]{Nogueras2018prep}
{Nogueras-Lara}, F., {Sch{\"o}del}, R., {Gallego-Calvente}, A.~T., {et~al.}
  2019{\natexlab{a}}, arXiv e-prints, arXiv:1908.10366

\bibitem[{{Nogueras-Lara} {et~al.}(2019{\natexlab{b}}){Nogueras-Lara},
  {Sch{\"o}del}, {Najarro}, {Gallego-Calvente}, {Gallego-Cano}, {Shahzamanian},
  \& {Neumayer}}]{Nogueras:2019b}
{Nogueras-Lara}, F., {Sch{\"o}del}, R., {Najarro}, F., {et~al.}
  2019{\natexlab{b}}, \aap, 630, L3

\bibitem[{{Parsa} {et~al.}(2017){Parsa}, {Eckart}, {Shahzamanian}, {Karas},
  {Zaja{\v{c}}ek}, {Zensus}, \& {Straubmeier}}]{parsa2017}
{Parsa}, M., {Eckart}, A., {Shahzamanian}, B., {et~al.} 2017, \apj, 845, 22

\bibitem[{{Pfuhl} {et~al.}(2011){Pfuhl}, {Fritz}, {Zilka}, {Maness},
  {Eisenhauer}, {Genzel}, {Gillessen}, {Ott}, {Dodds-Eden}, \&
  {Sternberg}}]{Pfuhl:2011uq}
{Pfuhl}, O., {Fritz}, T.~K., {Zilka}, M., {et~al.} 2011, \apj, 741, 108

\bibitem[{{Pietrinferni} {et~al.}(2004){Pietrinferni}, {Cassisi}, {Salaris}, \&
  {Castelli}}]{Pietrinferni:2004}
{Pietrinferni}, A., {Cassisi}, S., {Salaris}, M., \& {Castelli}, F. 2004, \apj,
  612, 168

\bibitem[{{Pietrinferni} {et~al.}(2006){Pietrinferni}, {Cassisi}, {Salaris}, \&
  {Castelli}}]{Pietrinferni:2006}
{Pietrinferni}, A., {Cassisi}, S., {Salaris}, M., \& {Castelli}, F. 2006, \apj,
  642, 797

\bibitem[{{Portail} {et~al.}(2017){Portail}, {Wegg}, {Gerhard}, \&
  {Ness}}]{Portail:2017}
{Portail}, M., {Wegg}, C., {Gerhard}, O., \& {Ness}, M. 2017, \mnras, 470, 1233

\bibitem[{{Portegies Zwart} {et~al.}(2002){Portegies Zwart}, {Makino},
  {McMillan}, \& {Hut}}]{Portegies-Zwart:2002fk}
{Portegies Zwart}, S.~F., {Makino}, J., {McMillan}, S.~L.~W., \& {Hut}, P.
  2002, \apj, 565, 265

\bibitem[{{Rui} {et~al.}(2019){Rui}, {Hosek}, {Lu}, {Clarkson}, {Anderson},
  {Morris}, \& {Ghez}}]{rui:2019}
{Rui}, N.~Z., {Hosek}, Matthew~W., J., {Lu}, J.~R., {et~al.} 2019, \apj, 877,
  37

\bibitem[{{Sch{\"o}del} {et~al.}(2007){Sch{\"o}del}, {Eckart}, {Alexander},
  {Merritt}, {Genzel}, {Sternberg}, {Meyer}, {Kul}, {Moultaka}, {Ott}, \&
  {Straubmeier}}]{schoedel2007}
{Sch{\"o}del}, R., {Eckart}, A., {Alexander}, T., {et~al.} 2007, \aap, 469, 125

\bibitem[{{Sch{\"o}del} {et~al.}(2014{\natexlab{a}}){Sch{\"o}del}, {Feldmeier},
  {Kunneriath}, {Stolovy}, {Neumayer}, {Amaro-Seoane}, \&
  {Nishiyama}}]{Schodel:2014fk}
{Sch{\"o}del}, R., {Feldmeier}, A., {Kunneriath}, D., {et~al.}
  2014{\natexlab{a}}, \aap, 566, A47

\bibitem[{{Sch{\"o}del} {et~al.}(2014{\natexlab{b}}){Sch{\"o}del}, {Feldmeier},
  {Neumayer}, {Meyer}, \& {Yelda}}]{schodel:2014bn}
{Sch{\"o}del}, R., {Feldmeier}, A., {Neumayer}, N., {Meyer}, L., \& {Yelda}, S.
  2014{\natexlab{b}}, Classical and Quantum Gravity, 31, 244007

\bibitem[{{Sch{\"o}del} {et~al.}(2009){Sch{\"o}del}, {Merritt}, \&
  {Eckart}}]{schoedel2009}
{Sch{\"o}del}, R., {Merritt}, D., \& {Eckart}, A. 2009, \aap, 502, 91

\bibitem[{{Sch{\"o}del} {et~al.}(2003){Sch{\"o}del}, {Ott}, {Genzel}, {Eckart},
  {Mouawad}, \& {Alexander}}]{Schoedel2003}
{Sch{\"o}del}, R., {Ott}, T., {Genzel}, R., {et~al.} 2003, \apj, 596, 1015

\bibitem[{{Sch{\"o}del} {et~al.}(2013){Sch{\"o}del}, {Yelda}, {Ghez}, {Girard},
  {Labadie}, {Rebolo}, {P{\'e}rez-Garrido}, \& {Morris}}]{Schodel:2013fk}
{Sch{\"o}del}, R., {Yelda}, S., {Ghez}, A., {et~al.} 2013, \mnras, 429, 1367

\bibitem[{{Sch{\"o}nrich} {et~al.}(2015){Sch{\"o}nrich}, {Aumer}, \&
  {Sale}}]{Schoenrich:2015}
{Sch{\"o}nrich}, R., {Aumer}, M., \& {Sale}, S.~E. 2015, \apjl, 812, L21

\bibitem[{{Stolte} {et~al.}(2008){Stolte}, {Ghez}, {Morris}, {Lu}, {Brandner},
  \& {Matthews}}]{Stolte:2008uq}
{Stolte}, A., {Ghez}, A.~M., {Morris}, M., {et~al.} 2008, \apj, 675, 1278

\bibitem[{{Stolte} {et~al.}(2014){Stolte}, {Hu{\ss}mann}, {Morris}, {Ghez},
  {Brandner}, {Lu}, {Clarkson}, {Habibi}, \& {Matthews}}]{Stolte:2014ys}
{Stolte}, A., {Hu{\ss}mann}, B., {Morris}, M.~R., {et~al.} 2014, \apj, 789, 115

\bibitem[{{Stolte} {et~al.}(2015){Stolte}, {Hu{\ss}mann}, {Olczak}, {Brandner},
  {Habibi}, {Ghez}, {Morris}, {Lu}, {Clarkson}, \& {Anderson}}]{stolte:2015}
{Stolte}, A., {Hu{\ss}mann}, B., {Olczak}, C., {et~al.} 2015, \aap, 578, A4

\bibitem[{{Wang} {et~al.}(2010){Wang}, {Dong}, {Cotera}, {Stolovy}, {Morris},
  {Lang}, {Muno}, {Schneider}, \& {Calzetti}}]{Wang:2010fk}
{Wang}, Q.~D., {Dong}, H., {Cotera}, A., {et~al.} 2010, \mnras, 402, 895

\bibitem[{{Yusef-Zadeh} {et~al.}(2009){Yusef-Zadeh}, {Bushouse}, {Wardle},
  {Heinke}, {Roberts}, {Dowell}, {Brunthaler}, {Reid}, {Martin}, {Marrone},
  {Porquet}, {Grosso}, {Dodds-Eden}, {Bower}, {Wiesemeyer}, {Miyazaki}, {Pal},
  {Gillessen}, {Goldwurm}, {Trap}, \& {Maness}}]{yusefzadeh2009}
{Yusef-Zadeh}, F., {Bushouse}, H., {Wardle}, M., {et~al.} 2009, \apj, 706, 348

\bibitem[{{Yusef-Zadeh} \& {Morris}(1987)}]{yusefzadeh1987}
{Yusef-Zadeh}, F. \& {Morris}, M. 1987, \apj, 320, 545

\bibitem[{{Zhao} {et~al.}(1993){Zhao}, {Desai}, {Goss}, \&
  {Yusef-Zadeh}}]{Zhao93}
{Zhao}, J.-H., {Desai}, K., {Goss}, W.~M., \& {Yusef-Zadeh}, F. 1993, \apj,
  418, 235

\end{thebibliography}

\newpage
 \begin{appendix}
 
 \section{}
In this section, we provide supplementary plots to the ones presented in the main body of the text.


    \begin{figure*}[]
     \begin{center}

        \subfigure{%
            \includegraphics[width=0.43\textwidth]{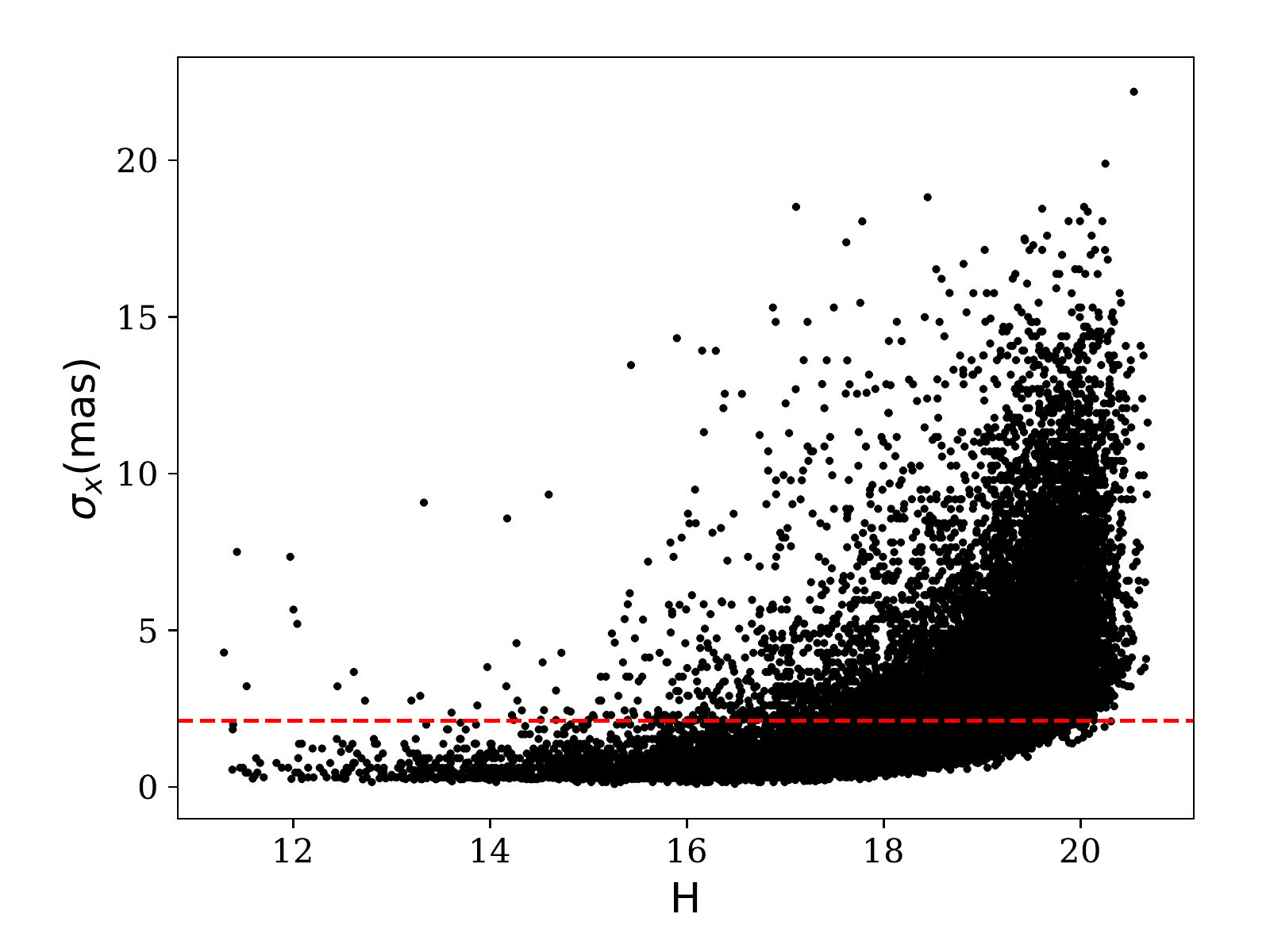}
        }
        \subfigure{%
          \includegraphics[width=0.43\textwidth]{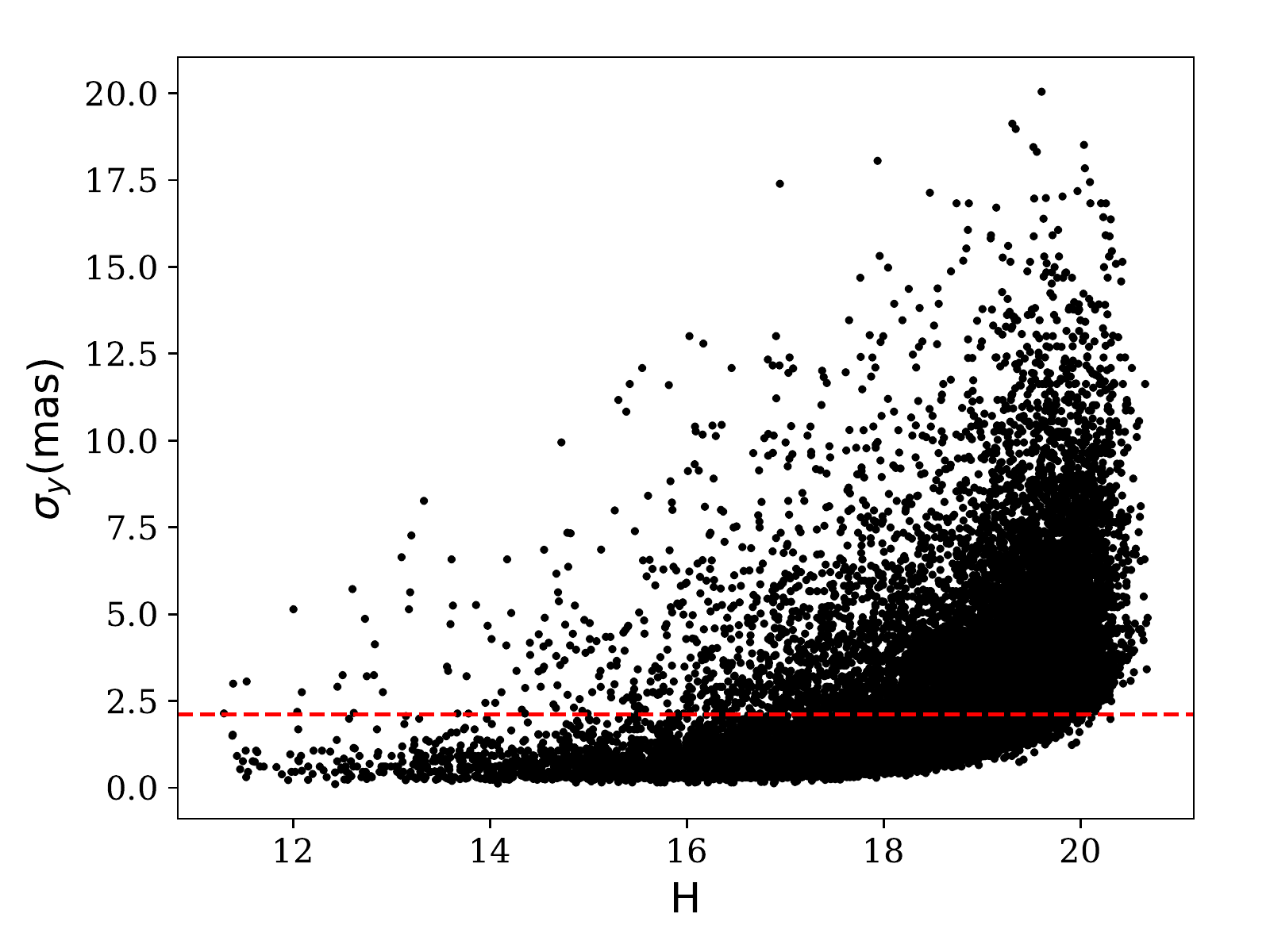}
        }\\

    \end{center}
    \caption{Relative astrometric uncertainties
 of chip \#1 of F19 for HAWK-I/VLT observations. For our proper motion analysis, we only considered the stars with position uncertainties less than 2~mas. 
    }
\label{fig:pos_uncer}
\end{figure*}

 

    \begin{figure*}[]
     \begin{center}

        \subfigure{%
            \includegraphics[width=0.43\textwidth]{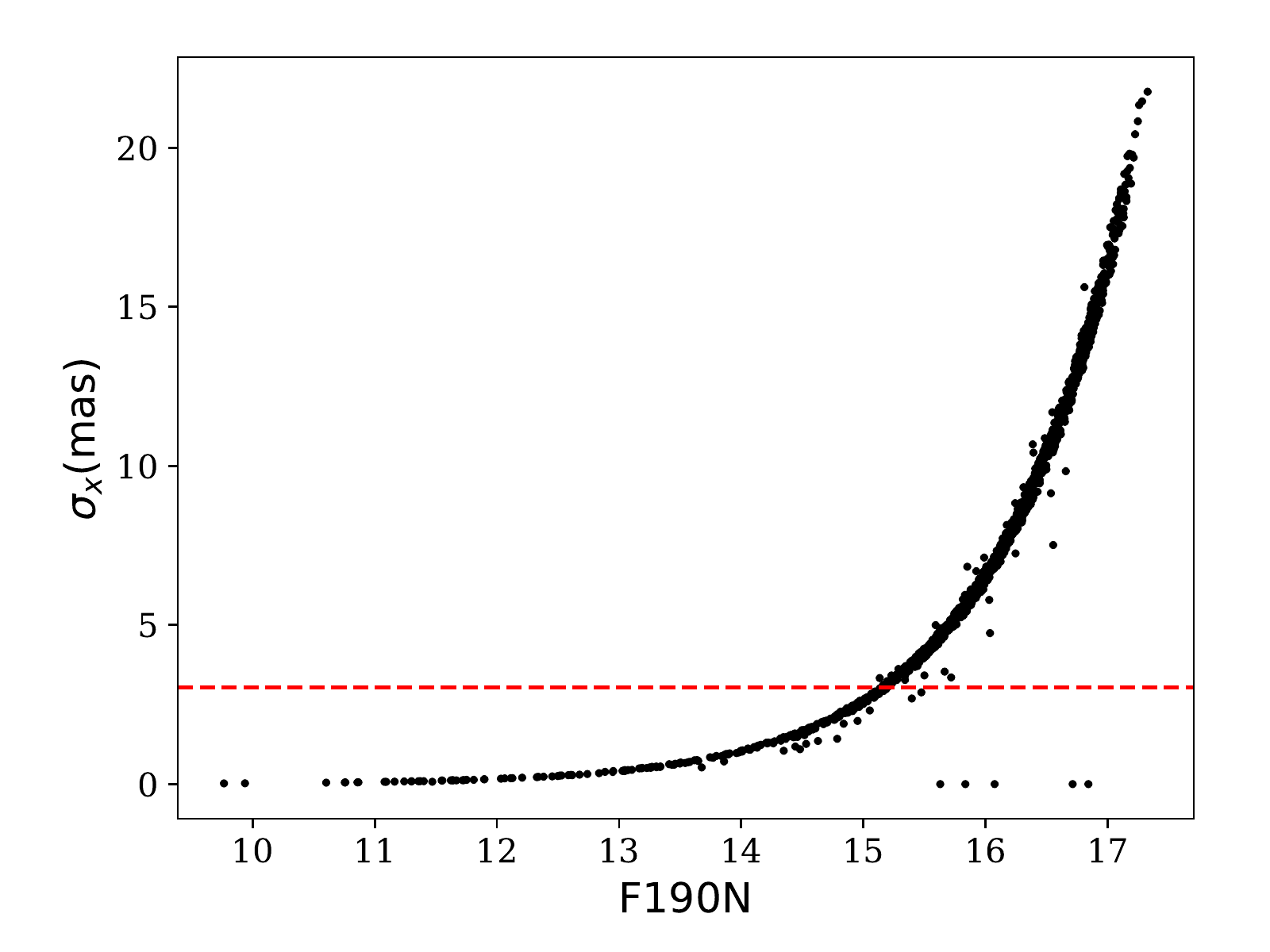}
        }
        \subfigure{%
          \includegraphics[width=0.43\textwidth]{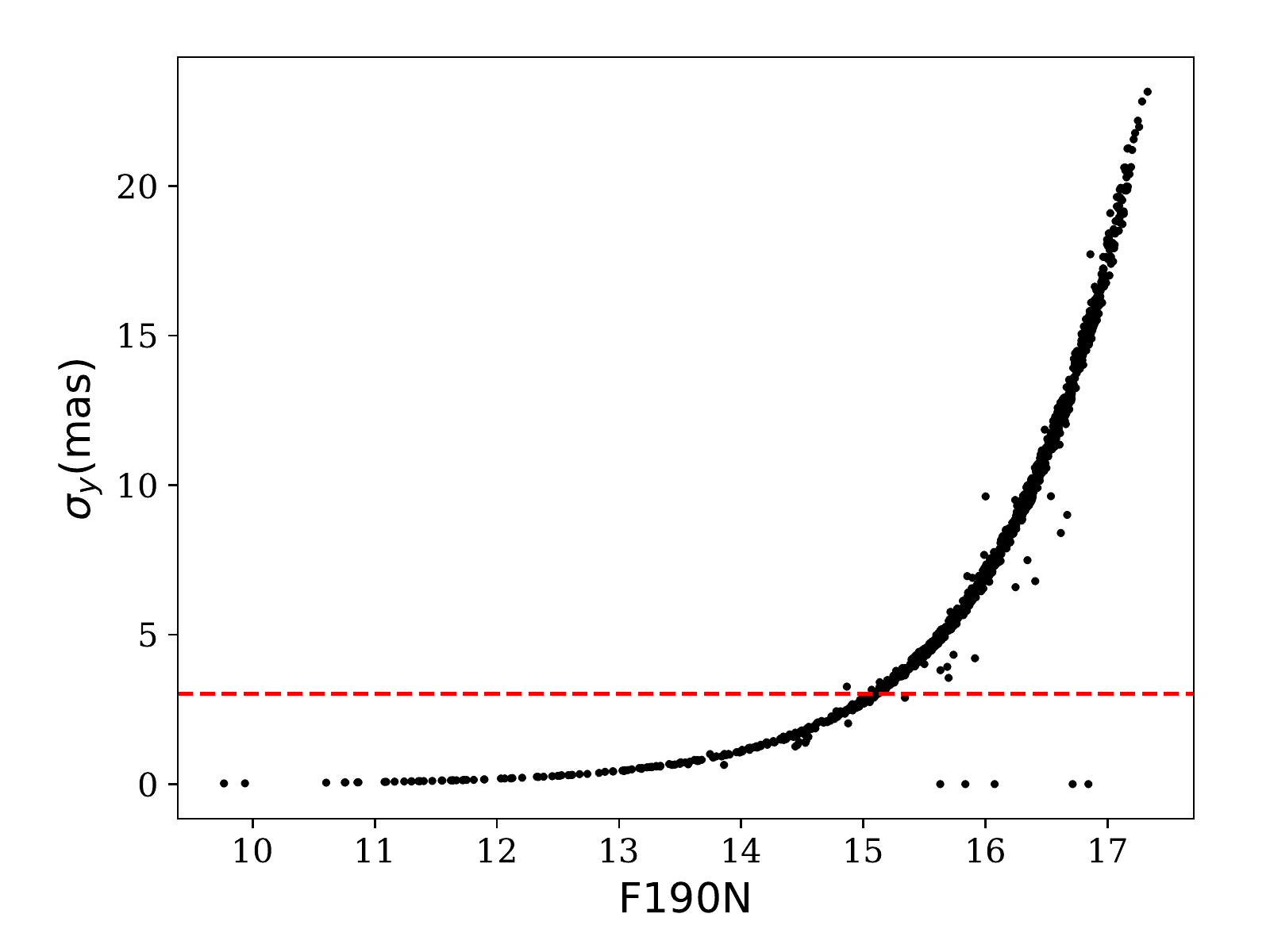}
        }\\

    \end{center}
    \caption{Relative astrometric uncertainties
 of one of the pointings of NICMOS/HST observations (that has an overlap with chip \#1 of F19 of HAWK-I/VLT). For our proper motion analysis, we only considered the stars with position uncertainties less than 3~mas. 
    }
\label{fig:pos_uncer_hst}
\end{figure*}


\begin{figure}[]
  \includegraphics[width=\columnwidth]{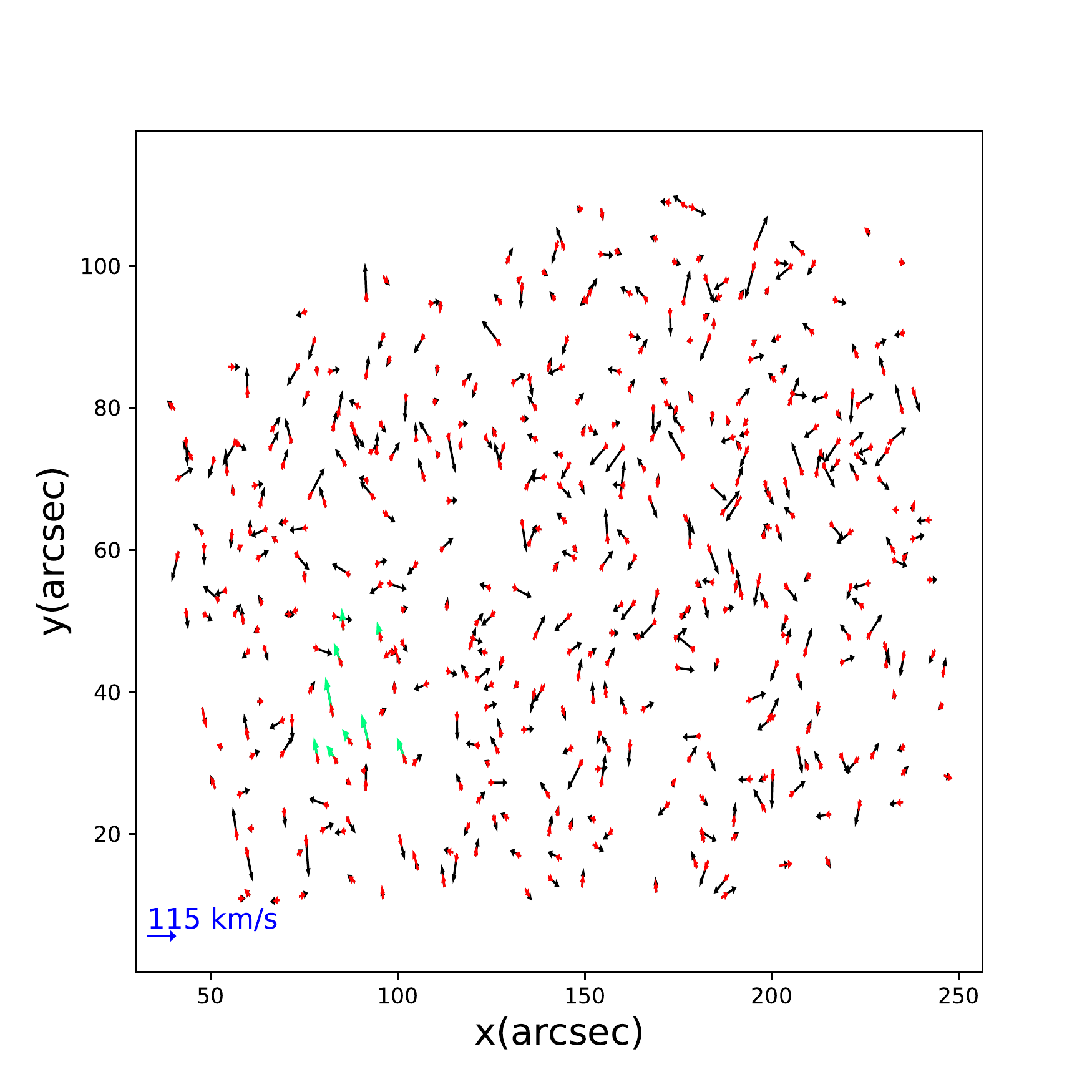}
  \caption{proper motion measurements of the stars in chip \#1 F19 with their uncertainties (the red arrows). The green arrows present the co-moving group of stars.}
  \label{fig:pm_arrow_comove}
\end{figure}


\section{Further tests of the cluster search analysis}

\begin{figure*}[]
      \begin{center}
    
        \subfigure{%
            \includegraphics[width=0.45\textwidth]{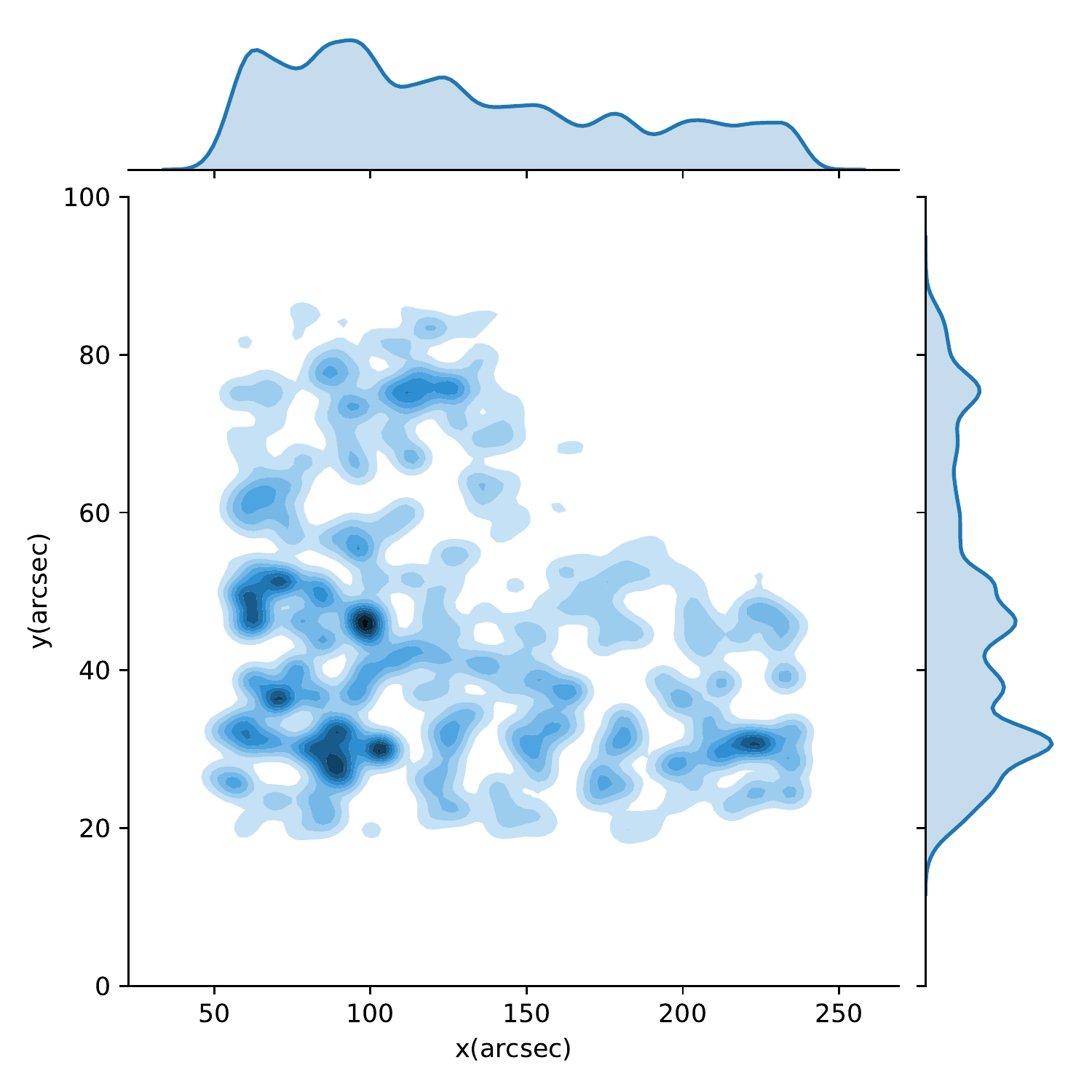}
        }
        \subfigure{%
          \includegraphics[width=0.45\textwidth]{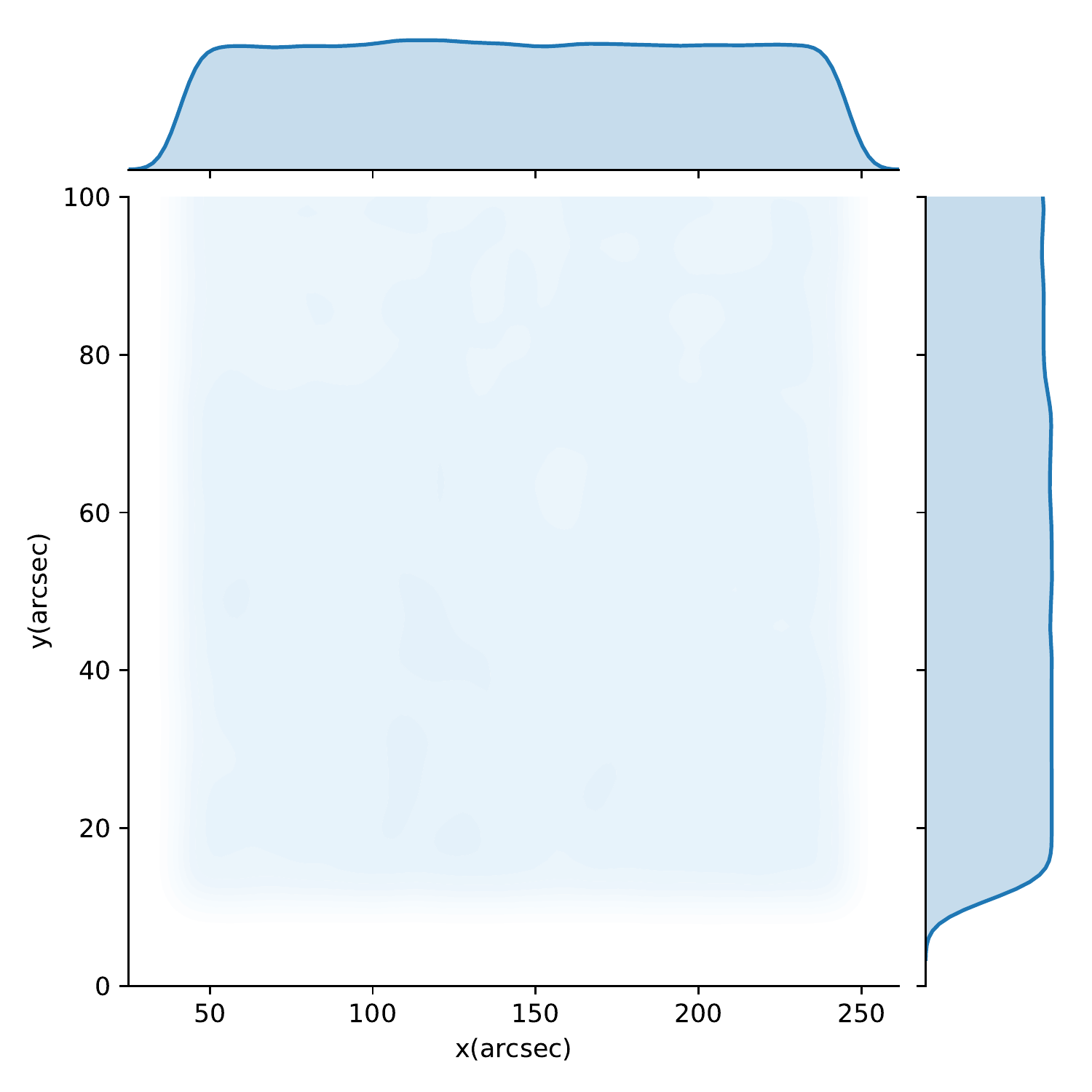}
        }\\

     \end{center}
     
    \caption{Left: Density map of \textit{$dis\underline{~}v_{min}$} for the stars of F19, chip \#1. The densest region is where the co-moving group of stars is located, and it also consists of the foreground stars since they move coherently with the rest of the stars in the group. Right: Density map of \textit{$dis\underline{~}v_{min}$} for the simulated stars.  }
\label{fig:jontplot_mc}
\end{figure*}

We checked if we are biased by the fact that we discovered the group of co-moving stars by eye, and if this group has the minimum $dis\underline{~}v$ and $dis\underline{~}p$ compared to the other groups of observed stars. We made a grid of the observational positions (x and y values) and their related velocities ($v_{x}$ and $v_{y}$) of chip \#1 of F19. By randomising the bin size in both directions of x and y, we performed a MC simulation and obtained the \textit{$dis\underline{~}v_{min}$} (minimum of \textit{$dis\underline{~}v$}) for 10000 times for the stars in each grid. Figure~\ref{fig:jontplot_mc}; left is an obtained density map of \textit{$dis\underline{~}v_{min}$} of our observational data, and it shows a peak at the position of our co-moving group of stars making a cluster. Therefore, our co-moving group of stars have the closest distance in both position and velocity space.

Furthermore, we used another time the MC simulation to statistically analyse the calculated \textit{$dis\underline{~}v$}.
We calculated 10000 times \textit{$dis\underline{~}v$} for the simulated stars with velocities from a Gaussian distribution prior with the median and $\sigma$ from the observed velocity distribution, and with positions from a uniform prior considering the lower and upper limits of the observed positions. We calculated for each grid of simulated stars (this time with fixed bin size) the \textit{$dis\underline{~}v_{min}$} from the simulated velocity distribution. The resulted density map of the simulated stars is presented in Fig.~\ref{fig:jontplot_mc}; right that does not show over-densities similar to Fig.~\ref{fig:jontplot_mc}; left, meaning that we could not produce the observed co-moving group of stars through the simulation.


\end{appendix}
\end{document}